\documentclass[usenatbib]{mnras}
\usepackage[T1]{fontenc}

\usepackage{graphicx}
\usepackage{amsmath}
\usepackage{amssymb}
\usepackage{verbatim}
\usepackage{orcidlink}

\title[FDM Halos with Attractive Self-Interactions]{An attractive model: simulating fuzzy dark matter with attractive self-interactions}

\author[Painter et al.]{
Connor A. Painter$^{1}$\orcidlink{0000-0002-3531-4806},
Michael Boylan-Kolchin$^{1}$\orcidlink{0000-0002-9604-343X},
Philip Mocz$^{2,3}$\orcidlink{0000-0001-6631-2566},
and Mark Vogelsberger$^{4}$\orcidlink{0000-0001-8593-7692}\\
$^{1}$Department of Astronomy, The University of Texas at Austin, 2515 Speedway, Stop C1400, Austin, TX 78712-1205, USA\\
$^{2}$Lawrence Livermore National Laboratory, 4 Ivy Lane, 7000 East Ave, Livermore, CA 94550, USA\\
$^{3}$Department of Astrophysical Sciences, Princeton University, 4 Ivy Lane, Princeton, NJ 08544, USA\\
$^{4}$Department of Physics, Kavli Institute for Astrophysics and Space Research, M.I.T., Cambridge, MA 02139, USA\\
}

\date{Accepted XXX. Received YYY; in original form ZZZ}
\pubyear{2023}

\begin{document}
\label{firstpage}
\pagerange{\pageref{firstpage}--\pageref{lastpage}}
\maketitle

\begin{abstract}
Fuzzy Dark Matter (FDM) comprised of ultralight ($m \sim 10^{-22}~\rm{eV}$) boson particles has received significant attention as a viable alternative to Cold Dark Matter (CDM), as it approximates CDM on large scales ($\gtrsim 1$~Mpc) while potentially resolving some of its small-scale problems via kiloparsec-scale quantum interference. However, the most basic FDM model, with one free parameter (the boson mass), is subject to a tension: small boson masses yield the desired cores of dwarf galaxies but underpredict structure in the Lyman-$\alpha$ forest, while large boson masses render FDM effectively identical to CDM. This \textit{Catch-22} problem may be alleviated by considering an axion-like particle with attractive particle self-interactions. We simulate an idealized FDM halo with self-interactions parameterized by an energy decay constant $f \sim 10^{15}~\rm{GeV}$ related to the axion symmetry-breaking conjectured to solve the strong-CP problem in particle physics. We observe solitons, a hallmark of FDM, condensing within a broader halo envelope, and find that the density profile and soliton mass depend on self-interaction strength. We propose generalized formulae to extend those from previous works to include self-interactions. We also investigate a critical mass threshold predicted for strong interactions at which the soliton collapses into a compact, unresolved state. We find that the collapse happens quickly and its effects are initially contained to the central region of the halo.
\end{abstract}





\section{Introduction}
\label{sec:intro}
The particle nature of cosmological dark matter is still one of the most pressing unknowns in modern astrophysics. For decades, Cold Dark Matter (CDM) has prevailed as the leading theory, stating that dark matter particles are non-relativistic, collisionless, and dissipationless. CDM, as part of the $\Lambda$CDM paradigm, has reproduced observations of the Cosmic Microwave Background \citep{aghanim2020, alam2021} and large-scale structure remarkably well \citep{vogelsberger2014, vogelsberger2014a, vogelsberger2020, schaye2014, springel2017}. However, the simplest CDM simulations admit puzzling discrepancies with observations on the scale of dwarf galaxies \citep{bullock2017, delpopolo2017, sales2022}. Problems actively debated in the literature include missing satellites \citep{klypin1999, moore1999}, density profile cores versus cusps \citep{moore1994, flores1994, deblok2010}, dark matter halos ``too big to fail'' to produce stars \citep{boylan-kolchin2011, garrison-kimmel2014}, and overly diverse galaxy rotation curves \citep{oman2015}. Even though baryonic feedback has shown promise to remedy many of the inconsistencies when incorporated into $\Lambda$CDM simulations, the most commonly considered CDM particle candidates, Weakly Interacting Massive Particles (WIMPs) on the mass scale of GeV, have so far evaded discovery \citep{roszkowski2018}. Small-scale inconsistencies along with non-detections of plausible particle candidates have fueled a search for alternative models.

A popular alternative to CDM is dark matter in the form of ultra-light boson particles of mass $m \sim 10^{-22}~\rm{eV}$ \citep{hu2000, guzman2003, hui2017, mocz2019, burkert2020, niemeyer2020, hui2021}. This so-called Fuzzy Dark Matter (FDM) model approximates CDM on large scales \citep{widrow1993, kopp2017}, but small-scale structure is altered by a ``quantum pressure'' tensor in the momentum equation \citep{schive2014a}. The dark matter clusters under self-gravity with fluid-like properties, and dark waves generated on the de Broglie scale $\lambda_{\rm{dB}} \equiv \frac{h}{mv} \sim \rm{kpc}$ interfere to smooth over small-scale structure. This smoothing cuts off the dark matter power spectrum above a certain wavenumber \citep{hu2000}, offering a natural explanation for the missing satellites predicted by CDM-only simulations. FDM also naturally addresses the cusp-core discrepancy: halos are characterized by cored central structures called \textit{solitons} \citep{schive2014a, schive2014} enveloped by a broader NFW-like power law drop-off in density \citep{navarro1996, marsh2015, mocz2017}. Furthermore, ultra-light bosons are predicted to arise naturally in many string theory models \citep{svrcek2006} and their present-day energy density in the Universe could be comparable to the measured dark matter density \citep{arvanitaki2010, marsh2016, hui2017}. 

These convenient properties of the FDM model have generated excitement and substantial investigation in the literature. In recent years, for example, FDM has been simulated both in high resolution on cosmological scales to characterize structure formation \citep{woo2009, mocz2019, mocz2020, lague2021, lague2023, nori2021, may2021, schwabe2022, huang2023, dome2023, shen2024} and on scales of individual halos in idealized scenarios \citep{mocz2017, du2018, veltmaat2020, schwabe2020, li2021}. In the simplest FDM model, the shape of a cosmological soliton is related to its total mass and the boson mass \citep{schive2014a}, and some regions of parameter space are capable of matching observations of dwarf galaxies \citep{marsh2015, luu2020, safarzadeh2020}. For a recent mini-review of the achievements of FDM, see \cite{matos2023}.

However, the simple FDM model struggles to simultaneously reproduce the power spectrum of the Lyman-$\alpha$ forest and the core sizes of dwarf galaxies \citep{irsic2017, nori2019, dome2024}. Lower boson masses \citep[$m<1.1 \times 10^{-22}~\rm{eV}$ at $2\sigma$ C.L.,][]{marsh2015} are required to yield the desired cores of satellite galaxies, but higher masses \citep[$m>2.0 \times 10^{-21}~\rm{eV}$ at $2\,\sigma$ C.L.,][]{irsic2017} are required to predict adequate small-scale structure in the Lyman-$\alpha$ forest. Constraints from the two observations leave little to no overlap. There is some discussion \citep[e.g.,][]{elgamal2023} suggesting that ``vanilla'' FDM is still capable of fitting observations of both Lyman-$\alpha$ structure and dwarf galaxies, but this \textit{Catch-22} \citep{davies2020} is widely regarded to be a serious challenge. Other constraints on FDM models come from strong lensing \citep{shevchuk2023}, ultrafaint dwarfs \citep{hayashi2021, dalal2022}, dynamical friction \citep{foote2023}, and cosmology \citep{li2014, li2017}.

The \textit{Catch-22} may be alleviated by introducing a second degree of freedom through a scalar potential term that naturally arises if the FDM particle is an ultralight \textit{axion-like particle} \citep{arvanitaki2020, mocz2023}. Ultralight axions are natural outcomes of particle physics models that solve the longstanding ``strong CP problem'' in Quantum Chromodynamics (QCD) \citep{peccei1977, weinberg1978}. In the models we consider here, the axion-like particle would have a decay constant (or symmetry-breaking scale) $f \sim 10^{15}~\rm{GeV}$ present in an additional scalar potential term in its governing equations. This new potential term will instigate attractive interparticle self-interactions (SI). These interactions are extremely weak, with quartic coupling $m^2/f^2 \sim 10^{-92}$ for our fiducial values, but theoretical and numerical work suggest that they may have non-negligible impacts on cosmic structure at low redshift \citep{desjacques2018,mocz2023}. In particular, attractive self-interactions introduce a critical mass scale for FDM halos above which the soliton collapses into an extremely compact state \citep{chavanis2011, chavanis2011a}, a process which may bolster small-scale structure enough to match observations of the Lyman-$\alpha$ forest \citep{mocz2023}.

Recent simulations have begun to characterize the extent to which axion-like self-interactions change FDM predictions. \cite{amin2019} carried out cosmological simulations of FDM with attractive SI to examine soliton formation and gravitational clustering. \cite{chen2021} simulated isolated clusters of FDM halos with both attractive and repulsive self-interactions. \cite{glennon2021} simulated idealized solitons with attractive SI, verifying criteria for soliton collapse and quantifying changes in tidal stripping timescales. \cite{mocz2023} used cosmological simulations to gauge the extent to which weak attractive self-interactions enhance small-scale structure in the cosmic web. \cite{jain2023a} published an integrator for FDM systems with general self-interactions. Besides attractive self-interactions, other extensions of FDM explored in the literature include repulsive self-interactions \citep{dawoodbhoy2021,shapiro2022}, multi-field FDM \citep{luu2020, luu2024, eby2020, guo2021, huang2023}, mixed CDM and FDM \citep{schwabe2020, lague2023}, and vector dark matter, where FDM is a higher-spin field \citep{amin2022}.

The purpose of this work is to provide a careful analysis of the interior structure of FDM halos under attractive self-interactions in idealized simulations. We pay particular attention to the regime of weak SI in which the soliton is noticeably influenced, but not so much that it exceeds the critical mass and collapses. This work informs the analysis of future cosmological simulations of FDM with attractive self-interactions that seek to break the \textit{Catch-22}.

The rest of this paper is organized as follows. In Section~\ref{sec:eqn} we review the governing equations of fuzzy dark matter, modify them to include self-interactions, and solve them under spherical symmetry to formulate predictions for halo density profiles. In Section~\ref{sec:simulations}, we describe the idealized halo simulations we perform to study the effects of including self-interactions. In Section~\ref{sec:results}, we present analysis and trends between simulations with varying SI strengths. We contextualize and conclude the work in Sections~\ref{sec:discussion} and \ref{sec:conclusion}.


\section{Physical Equations}
\label{sec:eqn}

Self-interacting fuzzy dark matter is governed by the Gross-Pitaevskii-Poisson (GPP) equations,
\begin{align} 
    i \hbar \left( \frac{\partial}{\partial t} \right)\psi &= \left( -\frac{\hbar^2}{2m}\nabla^2 + mV - \frac{4\pi\hbar^2 a_s}{m^2}\rho \right) \psi \label{eq:gross-pitaesvkii-poission-1}\\
    \nabla^2 V &= 4\pi G(\rho - \bar{\rho}) \label{eq:gross-pitaesvkii-poission-2}
\end{align}
which are equivalent to the Schr\"odinger equation where the potential is the self-potential due to self-gravity, plus a non-linear \textit{attractive} self-interaction term. $\psi$ is the wavefunction that describes the dark matter, normalized so that the dark matter density $\rho$ is equal to $|\psi|^2$. $\bar{\rho}$ is the local mean dark matter density, and the $s$-scattering length $a_s$ quantifies the SI strength. It is related to axion symmetry-breaking scale $f$ by
\begin{equation} \label{eq:s-scattering-length}
    a_s = \frac{\hbar c^3 m}{32 \pi f^2}.
\end{equation}
Other studies follow the convention that $a_s<0$ for attractive self-interactions, in which the right-hand side of Eq. \ref{eq:s-scattering-length} and the last term in Eq. \ref{eq:gross-pitaesvkii-poission-1} would be negated. In our simulations, the GPP equations evolve an initial mass distribution into a single isolated dark matter halo with a central soliton.\\

\subsection{Generalized Density Profile}
\label{sec:generalized-density-profile}

Following \cite{lora2014}, the GPP equations can be solved numerically by assuming that $\psi$ is spherically symmetric,
\begin{equation*}
    \psi(r,t) = e^{-i\gamma t/\hbar} \phi(r).
\end{equation*}
where $\phi(r)$ is a positive, decreasing profile of an FDM soliton. Substituting into Eq. \ref{eq:gross-pitaesvkii-poission-1} and rearranging yields
\begin{align*}
    -\frac{\hbar^2}{2m}\frac{1}{r}\frac{\partial^2}{\partial r^2} (r\phi) &= \gamma\phi - mV\phi + \frac{4\pi\hbar^2 a_s}{m^2} \phi^3 \\
    \frac{1}{r} \frac{\partial^2}{\partial r^2} (rV) &= 4\pi G \phi^2
\end{align*}
Introducing dimensionless variables,
\begin{equation*}
    \begin{split}
        \hat\phi &= \frac{\sqrt{4\pi G} \hbar}{m c^2} \phi\\
        \hat r &= \frac{mc}{\hbar} r\\
        \hat \gamma &= \frac{1}{mc^2} \gamma\\
    \end{split}
    \quad\quad
    \begin{split}
        \hat V &= \frac{1}{c^2} V\\
        \hat t &= \frac{mc^2}{\hbar} t\\
        \hat a_s &= \frac{c^2}{Gm} a_s
    \end{split}
\end{equation*}
the GPP equations take a simpler form:
\begin{align} \label{eq:gross-pitaesvkii-poisson-dimensionless}
    \frac{d^2}{d\hat{r}^2} (\hat{r}\hat{\phi}) &= 2\hat{r}(\hat{V}-\hat{\gamma})\hat{\phi} - 2\hat{r}\hat{a}_s\hat{\phi}^3\\
    \frac{d^2}{d\hat{r}^2} (\hat{r}\hat{V}) &= \hat{r}\hat{\phi}^2.
\end{align}

For an FDM soliton, we impose $r=0$ boundary conditions $\partial_r \hat\phi=0$, $\partial_r \hat V=0$, and $\hat\phi=\hat\phi_c$. For chosen values of $\hat\phi_c$ and $\hat a_s$, there is a discrete number of $\hat\gamma$ values $\{ \hat\gamma_0, \hat\gamma_1, \hat\gamma_2, \hdots \}$ for which the solutions converge as $r \rightarrow \infty$. Each $\hat\gamma_i$ corresponds to a solution $\hat\phi_i(r)$ which has $i$ nodes. We are interested in the \textit{ground state} solution $i=0$, the unique solution with $\phi(r)>0$ at all radii and finite total mass.

We want to solve for the density profile $\rho(r) = |\phi(r)|^2$ at various SI strengths in hopes of developing an approximate functional form. The dimensionless density profile differs from the physical profile by some factor $\phi_0 / \phi_c$, where $\rho_0 = \phi_0^2$ is the physical central density of the FDM halo. The GPP equations admit a scaling relation that implies a family of solutions for any particular solution:
\begin{equation}
\label{eq:scaling-relation}
    \{ r, \phi, \rho, a_s \} \rightarrow
    \{ \epsilon^{-1} r, \epsilon^2 \phi, \epsilon^4 \rho, \epsilon^{-2} a_s \}
\end{equation}
In particular, $a_s$ is a scale-\textit{dependent} quantity; however, the density profile fitting formula must ultimately be scale-\textit{independent}. 
To create a suitable self-interaction strength parameter, let $\hat\phi_c = 1$ and $\epsilon = \phi_0/\phi_c$. Define the scale-free parameter $\beta$ as
\begin{equation}
    \beta \equiv \epsilon^2 \hat a_s = \frac{\sqrt{4\pi G}\hbar}{mc^2}\rho_0^{1/2} \frac{c^2}{Gm} a_s
\end{equation}
In terms of $f$ and fiducial values,
\begin{equation}
\label{eq:beta}
    \beta = 0.238 \left( \frac{\rho_0}{10^{10}~M_{\odot}\rm{kpc}^{-3}} \right)^{1/2} m_{22}^{-1} f_{15}^{-2}
\end{equation}
where $m_{22}=m/(10^{-22}~\rm{eV})$ and $f_{15}=f/(10^{15}~\rm{GeV})$. This measure of SI strength is somewhat arbitrary (in the sense that we could have chosen any $\hat\phi_c$), but can be conceptualized as the dimensionless $s$-scattering length $\hat a_s$, scaled according to Eq. \ref{eq:scaling-relation} from the physical central density $\rho_0$ to $\hat\phi_c=1$.

In the non-interacting case ($a_s=0$), the density profile of a soliton is well-fit by the single-parameter formula 
\begin{equation} \label{eq:density-profile-no-SI}
    \rho_{\rm{sol}}(r) = \rho_0\left[ 1 + 0.091\left( \frac{r}{r_c} \right)^2 \right]^{-8}
\end{equation}
\citep{schive2014a}, where the core radius $r_c$ is related to the central density $\rho_0$ by
\begin{equation}
\label{eq:rho0-r_c-relation}
    \rho_0 = 1.9 \times 10^7 \left( \frac{10^{-22}\ \text{eV}}{m} \right)^2 \left( \frac{\text{kpc}}{r_c} \right)^4 \frac{M_{\odot}}{\text{kpc}^3}.
\end{equation}

\begin{figure}
    \centering
    \includegraphics[width=\columnwidth]{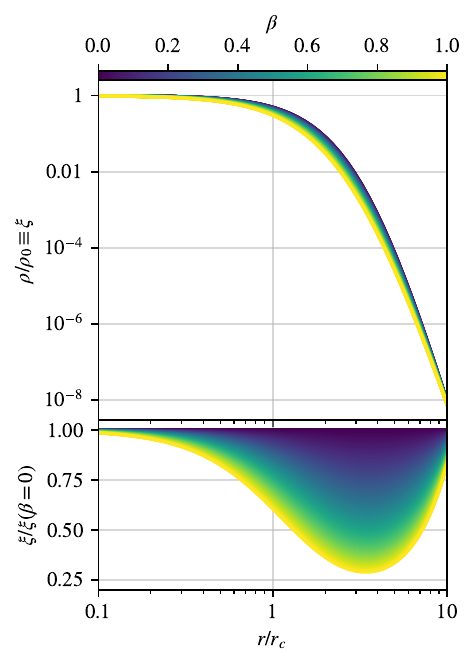}
    \caption{
    \textit{Upper}: Density profiles of FDM solitons with attractive self-interactions of varying strengths, normalized by the central density and core radius, as given by Eq. \ref{eq:density-profile-SI}. Self-interactions tend to shallow out the profile shape relative to the non-interacting case, decreasing the density at all radii within $r \lesssim 10\,r_c$.
    \textit{Lower}: The fractional change in the density profile relative to the non-interacting case $\beta=0$. The greatest fractional decreases in density occur around $r \sim 3.5\,r_c$
    }
    \label{fig:soliton-profile-beta-dependence}
\end{figure}

\begin{figure*}
    \centering
    \includegraphics[width=\textwidth]{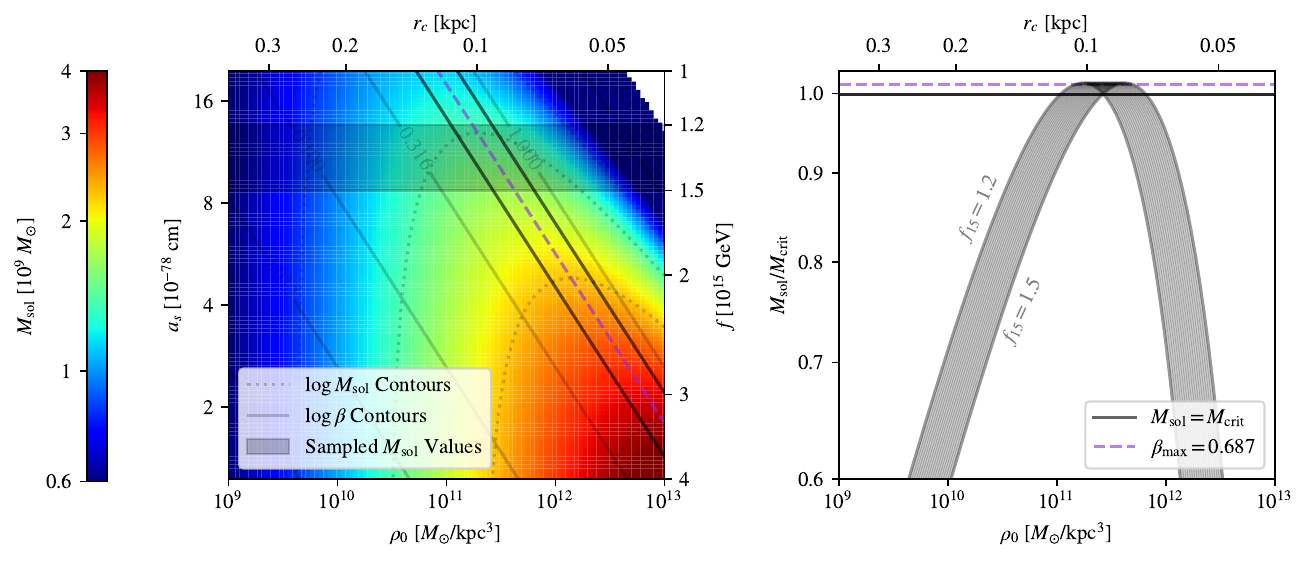}
    \caption{
    \textit{Left}: Heatmap of soliton mass $M_{\rm{sol}}$ for a range of central densities $\rho_0$ and self-interaction strengths $a_s$. Conversions to $r_c$ and $f$ are provided on the opposite axes. Overlaid are contours equally spaced in $\log M_{\rm{sol}}$ (dotted curves) and contours equally spaced in $\log\beta$ (solid curves). The two contours bolded black ($\beta=0.55,0.84$) indicate $M_{\rm{sol}}=M_{\rm{crit}}$ and $\beta_{\rm{max}}=0.687$ (purple dashed line) delineates the maximum predicted mass $M_{\rm{max}}=1.01M_{\rm{crit}}$ for any given $a_s$.
    \textit{Right}: $M_{\rm{sol}}$, normalized by the critical mass $M_{\rm{crit}}$, for every point in the shaded parameter space on the left panel. Bounding curves are labeled by $f_{15} \equiv f/(10^{15}~\rm{GeV})$. For any attractive SI strength, there exists a critical range of $\rho_0$ values in which the soliton is expected to collapse.
    }
    \label{fig:soliton-mass-grid}
\end{figure*}

In the general case ($a_s \ne 0$), we find that the ground state solutions of Eq. \ref{eq:gross-pitaesvkii-poission-1} are well-fit by a simple one-parameter extension to Eq. \ref{eq:density-profile-no-SI},
\begin{equation}
\label{eq:density-profile-SI}
    \rho_{\rm{sol}}(r) = \rho_0 \left[ 1 + 0.091\,a^2\left( \frac{r}{a r_c} \right)^{2-\beta/b} \right]^{-8}
\end{equation}
where $a=11.2$ and $b=4.2$ are best-fit constants to numerical solutions. Figure \ref{fig:soliton-profile-beta-dependence} plots how Eq. \ref{eq:density-profile-SI} predicts that the soliton profile will change with $\beta$, relative to the non-interacting case $\beta=0$. Based on this theoretical analysis, solitons with the same central density but stronger attractive self-interactions are expected to have \textit{shallower} density profiles, with matter redistributed out of the interior and toward the outskirts. This fitting formula is restricted in domain to $\beta < 2\,b$, but the SI strengths in this study are all well within this threshold. In fact, as we will show in Section \ref{sec:critical-mass}, solitons are likely to become unstable well before approaching $\beta=2\,b$. We validate the accuracy of Eq. \ref{eq:density-profile-SI} over the relevant range of SI strengths in Appendix \ref{sec:accuracy-appendix}.

The self-interaction strength adds an additional degree of freedom to the soliton density profile, but in a physical model, it is a set universal constant (like the boson mass), while the central density parameter $\rho_0$ varies from halo to halo. In other words, after $a_s$ and $m$ parameterize the axion-like particle, $\rho_0$ characterizes the individual halo. So, $a_s$ will vary between our simulations, but once it is set, $\rho_0$ is the only free parameter in Eq. \ref{eq:density-profile-SI}.

\subsection{Critical Mass} \label{sec:critical-mass}

While Eq. \ref{eq:density-profile-SI} is valid for fitting solitons in some cases, it does not apply in strong SI regimes. Axion self-interactions introduce a critical mass threshold for FDM solitons \citep{chavanis2011, chavanis2016, chavanis2018, chavanis2011a} that has been confirmed and studied in simulations \citep[e.g.,][]{chen2021, glennon2021, jain2024},
\begin{align}
\label{eq:critical-soliton-mass}
    M_{\rm{crit}} &= \frac{1.012\hbar}{\sqrt{Gma_s}}\\
    &= 1.1 \times 10^9 \frac{f_{15}}{m_{22}} M_{\odot}.
\end{align}
If the mass of a soliton exceeds $M_{\rm{crit}}$, it collapses from a ``dilute'' state \citep{chavanis2011a} into a compact state \citep{braaten2016}. The collapse remnant may be a ``boson star'' supported by higher-order terms in the axion potential and may be accompanied by a burst of relativistic axions, sometimes called a ``bosenova'' \citep{eby2016, levkov2017}. Alternatively, the collapse may lead to a black hole \citep{helfer2017}.

We can define the regions of $\rho_0$--$a_s$ parameter space where the soliton is expected to be stable or unstable by integrating the density profile given by Eq. \ref{eq:density-profile-SI} to get a mass and comparing to Eq. \ref{eq:critical-soliton-mass}. In the zero SI case $\beta=0$, the integral can be computed analytically,
\begin{align}
\label{eq:soliton-mass}
    M_{\rm{sol}} &= \int_0^{\infty} 4\pi r^2 \rho_{\rm{sol}}(r)\mathrm{d}r \approx 11.6\rho_0 r_c^3\\
    &= 2.2\times 10^8 \left( \frac{10^{-22}\ \text{eV}}{m} \right)^2 \left( \frac{\text{kpc}}{r_c} \right) M_{\odot}
\end{align}
demonstrating the unique, well-studied property that more-massive FDM solitons are smaller. For other values of $\beta$, Eq. \ref{eq:density-profile-SI} can be integrated numerically. The left panel of Figure \ref{fig:soliton-mass-grid} shows a heat map of $M_{\rm{sol}}$ as a function of $\rho_0$ (or, equivalently, $r_c$) and $a_s$ (or $f$). Contours equally spaced in $\log\beta$ are shown as faded gray lines. For any constant value of $\rho_0$, stronger self-interactions \textit{decrease} the soliton mass as compared to its non-interacting counterpart. Fixing $a_s$, increasing $\rho_0$ increases the soliton mass up to some maximum value before decreasing sharply. As shown in the right panel, the soliton mass exceeds the critical mass through a range of central densities and peaks at $M_{\rm{max}}=1.01M_{\rm{crit}}$, regardless of SI strength.

Using Eq. \ref{eq:density-profile-SI}, the soliton mass formula can be easily extended from Eq. \ref{eq:soliton-mass} with an additional multiplicative factor,
\begin{equation}
\label{eq:soliton-mass-SI}
    M_{\rm{sol}} = 11.6 \,\rho_0 r_c^3\, g(\beta)
\end{equation}
where $g$ is a smooth, monotonically decreasing function with $g(0)=1$. We do not attempt to characterize $g(\beta)$ analytically, but we provide details in Appendix \ref{sec:g-appendix}. Dividing Eq. \ref{eq:soliton-mass-SI} by Eq. \ref{eq:critical-soliton-mass} and using the definition of $\beta$, it can be shown that the ratio of the soliton mass to the critical mass is simply a function of $\beta$,
\begin{equation}
\label{eq:soliton-mass-ratio-beta-correspondence}
    \left( \frac{M_{\rm{sol}}}{M_{\rm{crit}}} \right)^2 = 3.8 \,\beta g(\beta)^2
\end{equation}
Thus, the maximum of $M_{\rm{sol}}/M_{\rm{crit}}$ occurs at the maximum of $\beta g(\beta)^2$, which is determined to be approximately
\begin{equation}
\label{eq:beta-maximizer}
    \beta_{\rm{max}} = 0.687.
\end{equation}
If the best-fit $\beta$ were to surpass this value at some point in a simulation, the soliton mass as calculated by integrating Eq. \ref{eq:density-profile-SI} would start to \textit{decrease}. More broadly, $M_{\rm{sol}} > M_{\rm{crit}}$ within the range $0.55 < \beta < 0.84$. In a simulated FDM soliton, $\beta$ may slowly increase with $\rho_0$ and surpass the lower bound, $\beta_{\rm{crit}}=0.55$, at which point it is expected to undergo the phase transition. This critical value of $\beta$ corresponds to a central density
\begin{equation}
\label{eq:rho-crit}
    \rho_{\rm{crit}} = 5.3 \times 10^{10} \, m_{22}^2 f_{15}^4~\frac{M_{\odot}}{\rm{kpc}^3}
\end{equation}
or a core radius
\begin{equation}
\label{eq:r_c-crit}
    r_{\rm{crit}} = 0.138 \,m_{22}^{-1} f_{15}^{-1}~\rm{kpc}
\end{equation}
These values of $\beta_{\rm{crit}}$, $\rho_{\rm{crit}}$, and $r_{\rm{crit}}$ will serve as reference values in our simulation analysis.

Beyond $M_{\rm{crit}}$, we include a higher-order relativistic correction to Eq. \ref{eq:gross-pitaesvkii-poission-1} that is non-negligible at very high densities,
\begin{equation}
\label{eq:gross-pitaesvkii-poisson-corrected}
    i \hbar \left( \frac{\partial}{\partial t} \right)\psi = \left( -\frac{\hbar^2}{2m}\nabla^2 + mV - \frac{4\pi\hbar^2 a_s}{m^2}\rho + \frac{32\pi\hbar^4 a_s^2}{3m^5 c^2}\rho^2 \right) \psi
\end{equation}
If a collapse occurred and the central density skyrocketed, this added term would contribute a \textit{repulsive} particle interaction and a stabilizing positive pressure,
\begin{equation}
\label{eq:repulsive-pressure}
    P_4 = \frac{64\pi^2 a_s^2 \hbar^4}{9m^6 c^2}\rho^3,
\end{equation}
This pressure would counterbalance the destabilizing pressure from the attractive term,
\begin{equation}
\label{eq:attractive-pressure}
    P_2 = -\frac{2\pi a_s \hbar^2}{m^3}\rho^2,
\end{equation}
at a very high density,
\begin{equation}
\label{eq:equilibrium-density}
    \rho_{\rm{eq}} = \frac{9c^2m^3}{32 a_s\hbar^2\pi},
\end{equation}
and above this density, $P_4$ dominates. The post-collapse equilibrium density---the density of a compact soliton---is much higher than the density of a dilute soliton; for fiducial values of $m$ and $f$, the critical central density, given by Eq. \ref{eq:rho-crit}, is $3.3\times10^{10}\,M_{\odot}\rm{kpc}^{-3}$, while the equilibrium density is $3.1\times10^{17}\,M_{\odot}\rm{kpc}^{-3}$, nearly 7 orders of magnitude higher. However, it is unclear that the Newtonian limit is a good approximation in this post-collapse regime. A full general relativistic treatment may be required and the collapse could result in a black hole.

\subsection{Fitting Algorithm}
\label{sec:fitting-algorithm}

The density field of any simulation snapshot can be decomposed into a spherically-averaged radial profile about the soliton center. For this work, we compute the density $\rho$ at some radius $r$ from the soliton center---here defined as the densest point---by sampling the grid at a large number of points within a thin spherical shell $[r-\frac{\Delta r}{2}, r+\frac{\Delta r}{2}]$. Each point is assigned a sub-pixel coordinate value within the shell, and the density at that point is the density of the grid cell that encloses it. $\rho(r)$ is then the average of the density at all the sampled locations.

We are interested in measuring how self-interactions alter the distribution of matter in the entire FDM halo, including the soliton and the outer envelope. We choose to fit the whole profile simultaneously by assuming that $\rho(r)$ is the sum of two component profiles, $\rho_{\rm{sol}}(r)$ and $\rho_{\rm{tail}}(r)$, where $\rho_{\rm{sol}}$ is given in Eq. \ref{eq:density-profile-SI} and $\rho_{\rm{tail}}$ is defined as
\begin{equation}
\label{eq:density-profile-tail}
    \rho_{\mathrm{tail}}(r) \equiv \rho_{0,\mathrm{tail}}\left[ 1 + \left( \frac{r}{r_{c,\mathrm{tail}}} \right)^2 \right]^{n_{\infty}/2}
\end{equation}
and $\rho_{0,\rm{tail}}$, $r_{c,\rm{tail}}$, and $n_{\infty}$ are all independent free parameters. At $r \gg r_{c,\rm{tail}}$, $\rho_{\rm{tail}}$ is approximately a power law with index $n_{\infty}$. At small radii, the tail component flattens to avoid contributing to the soliton core. A transition radius $r_{\rm{cutoff}}$ can be defined as the radius at which $\rho_{\rm{sol}}(r)$ drops below $\rho_{\rm{tail}}(r)$ by half a dex. To correct for the slight curvature inherent in Eq. \ref{eq:density-profile-tail}, we report the power-law slope at a radius $r_{\rm{midtail}}$ within the outer envelope,
\begin{equation}
\label{eq:n}
    n \equiv \frac{d\log\rho_{\mathrm{tail}}}{d\log r} = n_{\rm{\infty}}\frac{(r/r_{c,\mathrm{tail}})^2}{1 + (r/r_{c,\mathrm{tail}})^2} \Bigg|_{r=r_{\mathrm{midtail}}}
\end{equation}
where $r_{\rm{midtail}}$ is the log-mean of $r_{\rm{cutoff}}$ and $L/2$. In most snapshots, $n \approx n_{\infty}$.

In the full profile fit $\rho(r)=\rho_{\rm{sol}} + \rho_{\rm{tail}}$, there are four free parameters: $\rho_{0,\rm{sol}}$ from Eq. \ref{eq:density-profile-SI} (written there as $\rho_0$), $\rho_{0,\rm{tail}}$, $r_{c,\rm{tail}}$, and $n_{\infty}$ from Eq. \ref{eq:density-profile-tail}. Most important are the total central density $\rho_0 = \rho_{0,\rm{sol}}+\rho_{0,\rm{tail}}$ and the power-law index $n$ from Eq. \ref{eq:n}. We evaluate the goodness of fit using a sum of squares of differences metric,
\begin{equation}
\label{eq:goodness-of-fit}
    \delta^2 = \frac{1}{J}\sum_j (\log\rho(r_j) - \log\rho_{\rm{fit}}(r_j))^2,
\end{equation}
where $J$ is the number of sampled density values. Since we can use this formula in any arbitrary radial interval, we evaluate an overall goodness of fit $\delta^2$ as well as component evaluations $\delta_{\rm{sol}}^2$ and $\delta_{\rm{tail}}^2$.

\subsection{Conservation of Mass and Energy}

The system has conserved quantities, including its total mass,
\begin{equation}
\label{eq:total-mass}
    M = \int \rho\ \mathrm{d}^3 x,
\end{equation}
and its total energy,
\begin{align}
\label{eq:total-energy}
    E &= \int \frac{\hbar^2}{2m}(\nabla\sqrt{\rho})^2\ \mathrm{d}^3 x + \int \frac{1}{2}\rho v^2\ \mathrm{d}^3 x + \int \frac{1}{2}\rho V\ \mathrm{d}^3 x \\
    &= K_{\rho} + K_v + W,
\end{align}
where $K_{\rho}$ is the gradient energy due to the quantum pressure tensor, $K_v$ is the classical kinetic energy, $W$ is the potential energy, and $v = \mathrm{arg}(\psi)/m$ is the Madelung velocity. The total (quantum) kinetic energy is $K = K_{\rho}+K_v$. Like the density $\rho$, each of these energy components can be computed at every point in space and decomposed into a radial profile.


\section{Idealized Halo Simulations}
\label{sec:simulations}

We numerically simulate the merging and evolution of self-interacting fuzzy dark matter halos within a box with side length $L = 20\ \rm{kpc}$ and periodic boundary conditions. The only physical free parameter is the particle self-interaction strength, parameterized as $f_{15} \equiv f/(10^{15}\ \rm{GeV})$. We hold the dark matter particle mass fixed at its fiducial value of $m = 10^{-22}\ \rm{eV}$. Dark matter density is discretized onto grids with $N^3=200^3$ and $400^3$ cells for uniform spatial resolutions of $\Delta x = 0.10$ to $0.05~\rm{kpc}$. We vary the simulation length from $T = 4$ to $20$ Gyr. The box is initialized with a collection of subhalos (see \ref{sec:initial-conditions}) and evolved through time via the spectral method detailed in \cite{mocz2017}. The dynamics studied here do not include baryonic physics---the evolution is entirely described by Eqs. \ref{eq:gross-pitaesvkii-poisson-corrected} and \ref{eq:gross-pitaesvkii-poission-2}---but the code could be modified in future work to include gas dynamics and star formation, as in e.g. \cite{mocz2020}.

\subsection{Initial Conditions}
\label{sec:initial-conditions}

\begin{figure}
    \centering
    \includegraphics[width=\columnwidth]{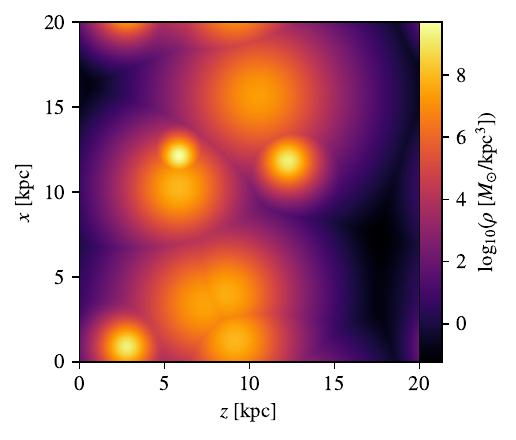}
    \caption{Density projection of the initial snapshot of our simulations, which is identical across all runs. A random number of bare solitons are strewn throughout the box with random masses and locations. Box properties are listed in Table \ref{table:initial-conditions}.}
    \label{fig:initial-snapshot}
\end{figure}

Regardless of the chosen SI strength, our simulations are always initialized with the same configuration of halos. The setup is determined by log-uniformly sampling between 4 and 32 halos. Each halo is initialized to be a spherically symmetric bare soliton with density profile given by Eq. \ref{eq:density-profile-no-SI} and core radius sampled as $r_c/\mathrm{kpc} \sim U(0.2,1)$. The halos are centered on randomly chosen locations in a box of side-length $L = 20\ \rm{kpc}$, then dark matter density $\rho$ is then initialized in an $N^3$ grid as the sum of contributions from each individual halo. The wavefunction $\psi$ is initialized and normalized as $\psi = \sqrt{\rho}$ and will be evolved by Eq. \ref{eq:gross-pitaesvkii-poisson-corrected}. No angular momentum is imparted to the system. A density projection of the randomized initial setup is provided in Figure \ref{fig:initial-snapshot} with some of its properties listed in Table \ref{table:initial-conditions}.

Initially, the solitonic subhalos are not given outer envelopes, as are expected from previous simulations, but the gravitational merging process is disruptive enough that the merged halo is not sensitive to this detail. The purpose of the chosen configuration is not to initialize realistic FDM halos from the start, but to set in motion a merging process that cascades turbulence throughout the box. The merged halo that forms will have the appropriate envelope even if the subhalos in the early configuration did not. 

\begin{table}
\centering
\begin{tabular}{ll}
\hline
\# of Subhalos & \;\;\,8 \\
$L$ & \;\;\,20 kpc \\
$M$ & \;\;\,$3.65 \times 10^9 M_{\odot}$ \\
$E$ & $-3.38 \times 10^{12} M_{\odot}\,(\rm{km/s})^2$\\
\hline
\end{tabular}
\caption{Initial conditions in our simulation box, regardless of resolution or self-interaction parameter. The random realization used in this paper is shown in Figure \ref{fig:initial-snapshot}.}
\label{table:initial-conditions}
\end{table}

\subsection{Numerical Evolution}
\label{sec:numerical-evolution}

To evolve and merge the halos, we numerically solve Eq. \ref{eq:gross-pitaesvkii-poisson-corrected} following the spectral method as developed and used in \cite{mocz2017}. The time-steps are decomposed into a kick-drift-kick leapfrog-like scheme, where each `kick' and `drift' are unitary operators acting on the wavefunction. The sequence is briefly reviewed here:

Once the density $\rho$ and wavefunction $\psi$ are discretized onto a grid of dimension $N^3$, the potential $V$ can be calculated by transforming to Fourier space and back,
\begin{equation}
\label{eq:potential}
    V = \mathrm{ifft}\left[ -\mathrm{fft}\left[ 4\pi G(\rho - \bar{\rho}) \right]/k^2 \right],
\end{equation}
where $\rm{fft}[\hdots]$ and $\rm{ifft}[\hdots]$ are the Fourier transform and inverse Fourier transform operators, respectively, and $k$ are the wavenumbers at the corresponding grid locations. The potential imparts a `kick' to the wavefunction, half a timestep forward,
\begin{equation}
\label{eq:kick}
    \psi \leftarrow \exp{\left[ -i(\Delta t/2)(m/\hbar)V \right]}\cdot\psi.
\end{equation}
This is followed by a full `drift' (kinetic) step in Fourier space:
\begin{align}
    \hat{\psi} &= \mathrm{fft}[\psi] \label{eq:drift1}\\
    \hat{\psi} &\leftarrow \exp{\left[ -i\Delta t (\hbar/m)k^2/2 \right]} \label{eq:drift2} \\
    \psi &\leftarrow \mathrm{ifft}[\hat{\psi}] \label{eq:drift3}
\end{align}
The timestep is completed with another `kick' step using Eq. \ref{eq:kick}, except that the interaction terms from Eq. \ref{eq:gross-pitaesvkii-poisson-corrected} are included in the potential,
\begin{equation}
    V \leftarrow V - \frac{4\pi\hbar^2 a_s}{m^3}|\psi|^2 + \frac{32\pi\hbar^4 a_s^2}{3m^6 c^2}|\psi|^4
\end{equation}
and the system is thus evolved from time $t$ to $t+\Delta t$.\\

The valid timestep criterion for stability and accuracy of our method, essentially a Courant–Friedrichs–Lewy (CFL) like condition, is that the unitary operators in Eqs. \ref{eq:kick} and \ref{eq:drift2} do not change the phase by more than $2\pi$ in each timestep. The timestep criterion of \cite{schwabe2016},
\begin{equation}
\label{eq:timestep-criterion}
    \Delta t \leq \max\left[ \frac{m}{6\hbar}(\Delta x)^2, \frac{h}{m\max|V|} \right]
\end{equation}
enforces this property, where $\Delta x = L/N$ is the grid spacing. Note that the timestep scales as $(\Delta x)^2$, which adds computational cost for high-resolution simulations. Unique to self-interacting fuzzy dark matter, if a soliton collapses into a very dense object at which the higher-order repulsive term from Eq. \ref{eq:gross-pitaesvkii-poisson-corrected} dominates, $\max|V|$ will grow very large, further suppressing the timestep.

A snapshot of the wavefunction is periodically stored to disk, along with the simulation parameters and current time. At any given snapshot, all relevant quantities can be derived from $\psi$ for subsequent analysis.


\section{Results}
\label{sec:results}

\begin{table*}
\begin{tabular}{|l|l|l|l|l|l|l|}
\hline
\# & 
$f_{15}$ & 
$N$ & 
$\Delta x$ {[}kpc{]} & 
$\Delta t$ {[}Gyr{]} & 
$T$ {[}Gyr{]} 
& Purpose \\ 
\hline

6 & 
\{$\infty,2,1.5,1.2,1.1,1.0$\} & 
400 & 
0.05 &
0.1 &
4 & 
Highest-resolution simulations.\\ 

4 & 
\{$\infty,2,1.5,1.2$\} & 
200 & 
0.10 & 
0.1 & 
20 & 
Long-term evolution.\\ 

24 & 
Evenly-spaced in $a_s$ in interval {[}$\infty$,1.2{]} & 
200 & 
0.10 & 
0.1 & 
4 & 
Quantifying trends with SI strength.\\ 
\hline
\end{tabular}
\caption{All simulations used in this paper, along with their purposes. For our initial conditions, the boundary between weak and strong interactions is between $f = 1.2$ and $1.1 \times 10^{15}$ GeV.}
\label{table:simulations-summary}
\end{table*}

\begin{figure*}
    \centering
    \includegraphics[trim={1cm 3cm 1cm 3cm}, width=\textwidth]{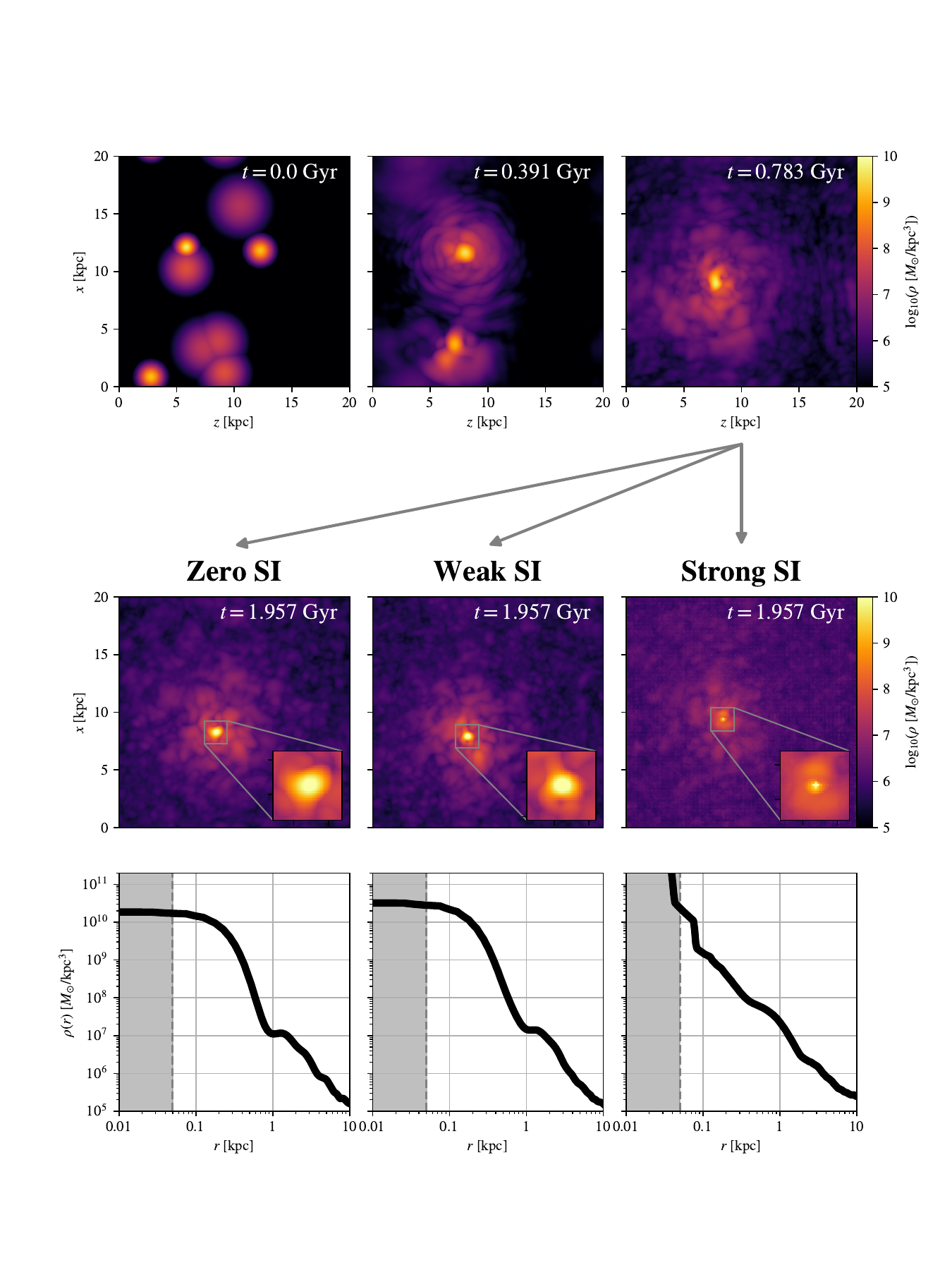}
    \caption{Evolution of halos in our simulations. Initially, the subhalos merge together in the same way (top panels), but the end state at $\sim 2$~Gyr depends on the strength of the attractive SI. The simulations diverge once the subhalos are fully merged and solitons condense in their centers (highlighted by the inset panels). As SI strength increases in the weak regime, the soliton becomes more compact while remaining in a dilute state. At some critical point that delineates the strong-SI regime, the self-interaction becomes strong enough to initiate a collapse into an unresolved compact soliton even though $M_{\rm{sol}} < M_{\rm{crit}}$.}
    \label{fig:evolution-branches}
\end{figure*}

In this work, our data consists of three sets of simulations, carried out as described in the previous section and detailed in Table \ref{table:simulations-summary}. Our highest-resolution simulations span a range of SI strengths between $f_{15}=\infty$ and $f_{15}=1.0$ with $N^3=400^3$ grid elements for a spatial resolution of $\Delta x=0.05~\rm{kpc}$. Our two suites of lower-resolution ($N=200$) simulations are designed to provide insight into the long-term evolution of the halo and the precise trends and changes when SI strength is varied.

Figure \ref{fig:evolution-branches} diagrams the general evolution of dark matter in our simulations. The top row of density projections depicts the early stages, before the subhalos have fully merged together. The dynamics at these times are largely invariant of SI strength in the ranges we probed. In all cases, the subhalos collide to produce a typical FDM halo with a soliton at its center, NFW-like outer envelope, and turbulent `granules' throughout the box. The simulations diverge after this point, represented by branches to three categories: non-interacting, weakly interacting, and strongly interacting cases.

In weakly interacting cases ($f_{15} = 2.0, 1.5, 1.2$), the soliton settles into a final state that depends on the interaction strength: stronger attractive interactions compactify the soliton to higher central densities and smaller radii. We examine dilute solitons in detail in Section \ref{sec:weakly-self-interacting-halos}.

In strongly interacting cases ($f_{15} = 1.1, 1.0$), the soliton collapses rapidly after some length of time that depends on the interaction strength. The uniform spatial resolution $\Delta x = 0.05\ \rm{kpc}$ is not enough to resolve the post-collapse compact objects. Compact solitons appear as a single high-density volume pixel which propagates numerical inaccuracies throughout the simulation box. The potential increases steeply and Eq. \ref{eq:timestep-criterion} enforces that integration proceeds much more slowly. Since further evolution is inaccurate and computationally expensive, we halt these simulations shortly after collapse (before $T = 4~\rm{Gyr}$). We discuss the dynamics of collapse and the properties of compact solitons in Section \ref{sec:strongly-self-interacting-halos}. 

\subsection{Weakly Self-Interacting Halos}
\label{sec:weakly-self-interacting-halos}

\subsubsection{Density Profiles}
\label{sec:density-profiles}

\begin{figure*}
    \centering
    \includegraphics[width=\textwidth]{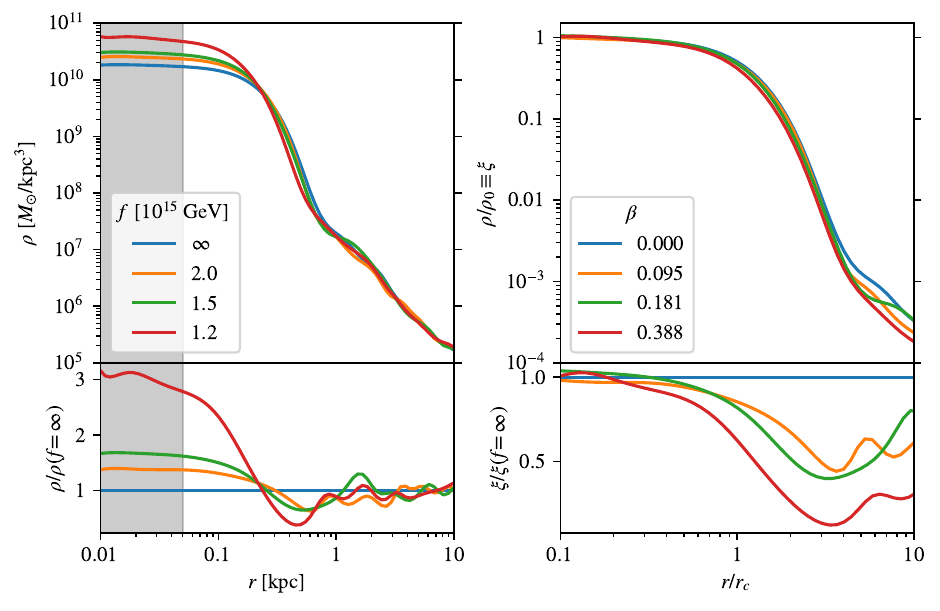}
    \caption{\textit{Left}: Measured density profiles at the end of the $N=400$, $f_{15}=\infty$ (blue), $2$ (orange), $1.5$ (green), and $1.2$ (red) simulations. Stronger self-interactions increase the central density and decrease the core radius, making the soliton more compact. The outer envelope is not affected by self-interactions, except that it surrounds a smaller soliton. \textit{Right}: Same profiles, normalized on both axes by $r_c$ and $\rho_0$ (from best-fit models). The zero SI case is well fit by Eq. \ref{eq:density-profile-no-SI}, but self-interactions introduce slight changes to the soliton shape that render Eq. \ref{eq:density-profile-SI} more accurate. Computed values of $\beta$ are included for comparison to Figure \ref{fig:soliton-profile-beta-dependence}, which shows the theoretical prediction.}
    \label{fig:density-profile-comparison}
\end{figure*}

\begin{figure}
    \centering
    \includegraphics[trim={0.2in 0 0.1in 0},  width=\columnwidth]{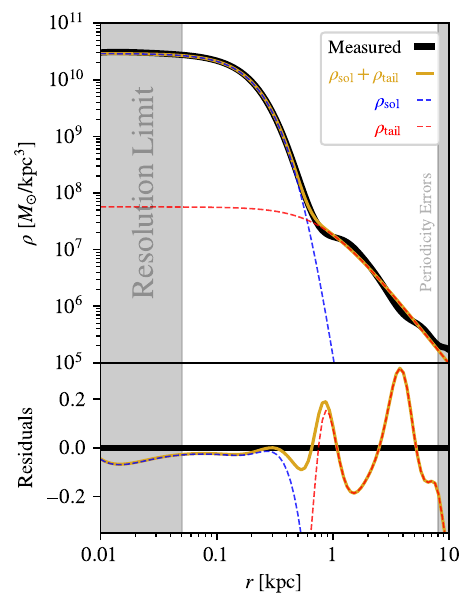}
    \caption{
    \textit{Upper}: Example density profile measured from the $f_{15}=1.5$ simulation (thick black curve) performed at the highest resolution, with best-fit model (gold curve) and fit components (red and blue dashed curves). The region interior to the grid cell length $\Delta x=0.05\,{\rm kpc}$ is shaded in gray; regions beyond $\sim 0.4\,L=8\,{\rm kpc}$ are subject to periodicity errors. Equation \ref{eq:density-profile-SI} fits the soliton profile very well, while the outer envelope is roughly a power law with turbulent fluctuations.
    \textit{Lower}: Ratio of best-fit components to the measured profile.}
    \label{fig:density-profile-example}
\end{figure}

Density profiles of self-interacting FDM halos with dilute solitons reproduce the key features FDM halos observed in previous simulations: a central soliton surrounded by a power-law tail. Figure \ref{fig:density-profile-comparison} plots the density profile measured at the end of the higher-resolution $f_{15} = \infty$ (blue), $2.0$ (orange), $1.5$ (green), and $1.2$ (red) simulations. In each case, the soliton is clearly visible protruding within the central kiloparsec. It has a cored center, with the density falling off at increasingly steep rates until some transition radius $r_{\rm{cutoff}}$. At $r>r_{\rm{cutoff}}$, the density profile assumes an NFW-like power-law shape with some fluctuations due to random interference granules. All self-interacting FDM halos with dilute solitons demonstrate these properties in our simulations.

Figure \ref{fig:density-profile-comparison} reveals that, by the end of the simulations, self-interactions have caused multiple changes to the shape of the central soliton. The left panel clearly suggests an important trend: \textit{stronger} self-interactions make the soliton \textit{more compact}. In other words, as we decrease $f_{15}$, we observe that the central density increases and the core radius shrinks. This makes intuitive sense: the soliton is the densest part of the halo in which the most particle-particle interactions will occur. Since that interaction is attractive, the soliton should compress into itself more. The degree of compression should be directly related to the SI strength, and this is reflected by the fact that the central densities are (inversely) sorted by $f_{15}$. By contrast, the outer envelope is unchanged by self-interactions; it has the same slope and amplitude at all radii exterior to $r_{\rm{cutoff}}$. For the smaller solitons in the self-interacting cases, the power law extrapolates inward to a new, smaller cutoff radius.

For the most part, the soliton shapes all resemble the zero-interaction case, but by normalizing $\rho$ by the central density $\rho_0$ and $r$ by the core radius $r_c$, as shown in the right panel, the minute variations from Eq. \ref{eq:density-profile-no-SI} are exposed. For a given $\rho_0$, the density of self-interacting solitons falls off slightly faster than is allowed by Eq. \ref{eq:density-profile-no-SI}, to the point that the $f_{15}=1.2$ soliton has $1/4$ the ``expected'' density near the cutoff radius. Fitting Eq. \ref{eq:density-profile-no-SI} to the soliton component of the red curve yields a relatively poor goodness-of-fit value of $\delta^2 = 0.0571$, while Eq. \ref{eq:density-profile-SI} successfully predicts these minute changes, boasting a best-fit $\delta^2 = 0.0054$. As mentioned in Section \ref{sec:generalized-density-profile}, both of these formulas have only one free parameter.

Figure \ref{fig:density-profile-example} isolates one of the density profiles in Figure \ref{fig:density-profile-comparison} ($f_{15}=1.5$, the green curve) and includes the theoretical best fit as computed by the algorithm in Section \ref{sec:fitting-algorithm}. The soliton and tail components of the fit are shown in red and blue, and their sum is plotted in gold. The log residuals are shown in the bottom panel. The soliton is fit very well by Eq. \ref{eq:density-profile-SI}, with goodness of fit $\delta_{\rm{sol}}^2 = 0.0085$. The (approximate) power law is a good fit for the tail, though natural random fluctuations increase the residuals to $\delta_{\rm{tail}}^2=0.024$. All other snapshots of halos with dilute solitons exhibit similarly excellent fits.

\subsubsection{Halo Evolution}

\begin{figure*}
    \centering
    \includegraphics[width=\textwidth]{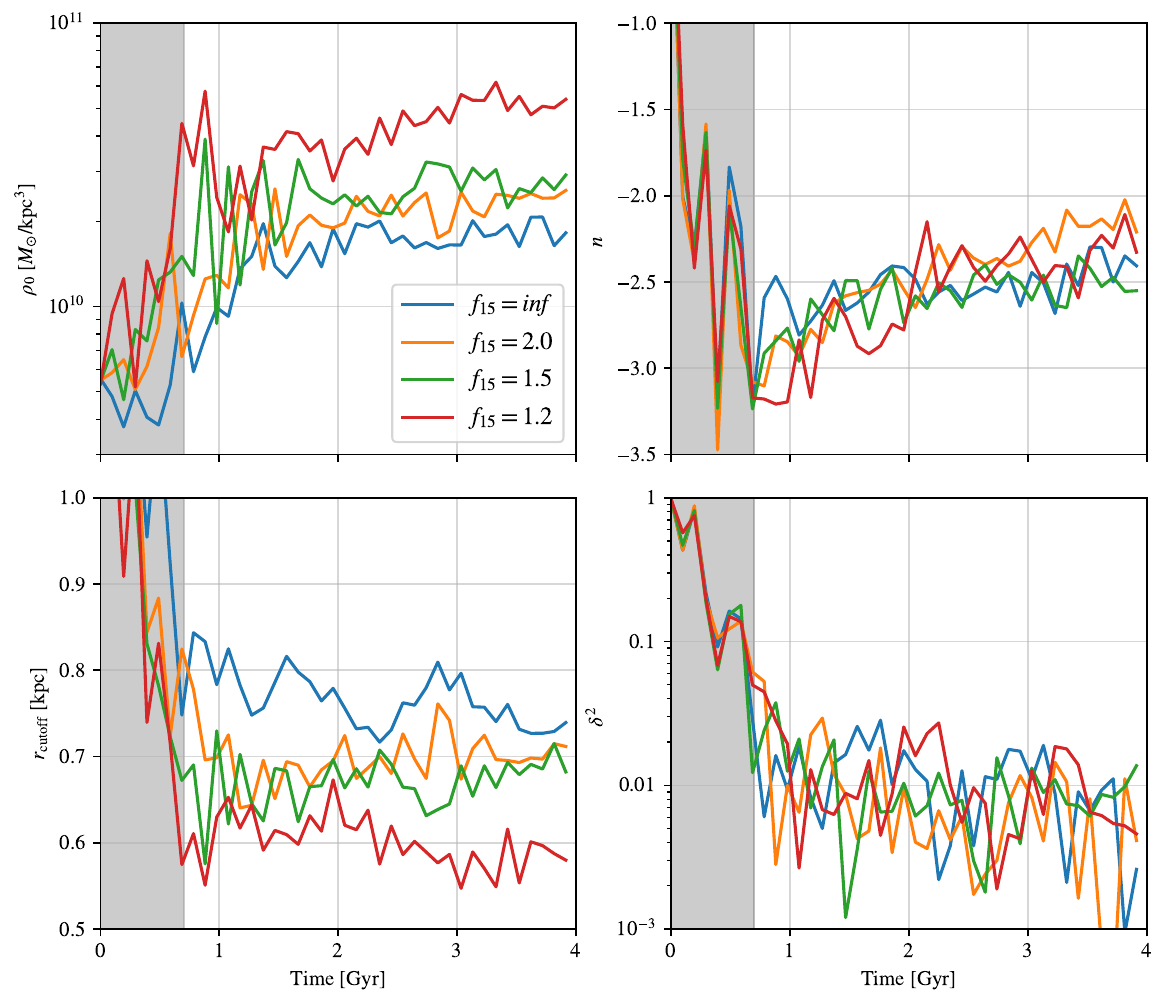}
    \caption{Best-fit quantities over time for weakly self-interacting FDM halos. The time before all subhalos have fully merged is indicated by the shaded area.
    \textit{Upper left}: The soliton central density $\rho_0$ exhibits a clear pattern across the four SI strengths by $t \approx 1.5~\rm{Gyr}$. Stronger self-interactions increase $\rho_0$ and thus decrease $r_c$.
    \textit{Upper right}: The power-law index $n$ of the outer envelope is independent of $f_{15}$ and increases slowly over the course of the simulation.
    \textit{Lower left}: The soliton cutoff radius $r_{\rm{cutoff}}$ decreases with stronger self-interactions, scaling in the same way as $r_c$.
    \textit{Lower right}: The goodness of fit does not depend on $f_{15}$, oscillating around $\delta^2=10^{-2}$. Our fitting algorithm therefore appears to be valid across the weak SI regime.
    }
    \label{fig:high-res-evolution}
\end{figure*}

\begin{figure*}
    \centering
    \includegraphics[width=\textwidth]{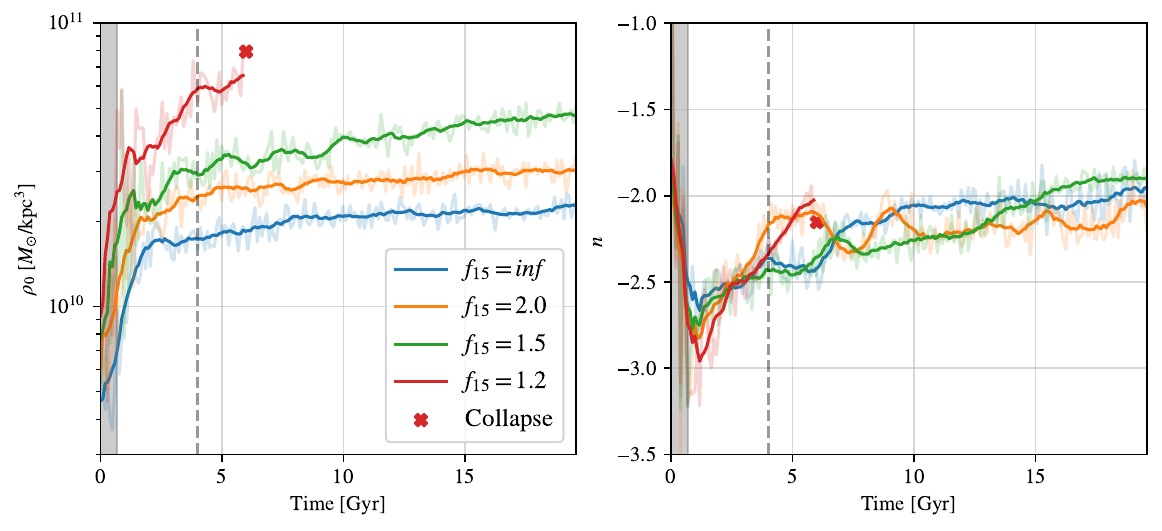}
    \caption{Extended evolution of the halos in Figure \ref{fig:high-res-evolution} using lower-resolution simulations ($N=200$). These simulations reproduce the trends observed through the first 4 Gyr in Figure \ref{fig:high-res-evolution} and continue them for an additional 16 Gyr. The soliton in the $f_{15}=1.2$ simulation is observed to collapse at $t = 5.87~\rm{Gyr}$, when it reaches its highest-yet central density of $\rho_0 = 8.03 \times 10^{10}~M_{\odot}/\rm{kpc}^3$. In the $f_{15}=\infty,2$, and $1.5$ simulations, both $\rho_0$ and $n$ continue their secular increase over long timescales, potentially reaching an equilibrium state in each case case.}
    \label{fig:long-term-evolution}
\end{figure*}

The density profile fitting algorithm exemplified in Figure \ref{fig:density-profile-example} can be applied to all snapshots outputted throughout each simulation to measure the essential halo quantities over time. Figure \ref{fig:high-res-evolution} shows the evolution of $\rho_0$, $n$, $r_{\rm{cutoff}}$, and $\delta^2$ derived from density profile fits for each of the higher-resolution $f_{15}=\infty, 2.0, 1.5$, and $1.2$ simulations. The initial subhalos do not fully merge together until $t_{\rm{merge}} \approx 0.7\ \rm{Gyr}$, indicated by a gray shaded region, at which point an FDM halo forms with a dilute soliton and outer envelope.

Even though each simulation begins with the same initial density field (projected in Figure \ref{fig:initial-snapshot}), the upper left panel shows that the best-fit soliton central densities sort themselves by $f_{15}$ after $t \approx 1.5\ \rm{Gyr}$. This reflects observations of the raw density profiles in Figure \ref{fig:density-profile-comparison}. Turbulence in the box means the evolution of $\rho_0$ is noisy, but in all cases a secular increase is observed after $t_{\rm{merge}}$. Since $M_{\rm{sol}} \propto \rho_0^{1/4}$, this may be interpreted as a slow accretion of the outer envelope over time.

The upper right panel plots the best-fit power-law index to the density profile tail over time. Immediately after $t_{\rm{merge}}$, the slope is $n \approx -3.0$, but over time it slowly shallows to $n \approx -2.4$ by $t = 4$, independent of the SI strength.

After $t_{\rm{merge}}$, $r_{\rm{cutoff}}$ stabilizes at a radius dependent on $f_{15}$, similar to the soliton central density. In fact, $r_{\rm{cutoff}}/r_c$ is invariant of $f_{15}$, so $r_{\rm{cutoff}}$ and $r_c$ scale with interaction strength in the same way. Unlike the core radius (which decreases over time since it is inversely related to $\rho_0$), the cutoff radius is not observed to change significantly over time. This means that later in its evolution, the soliton extends farther out relative to its core radius.

The goodness of fit appears to be independent of $f_{15}$, as the fits in each simulation only slightly deviate about $\delta^2 = 0.01$. This provides evidence that our $\rho_{\rm{sol}}+\rho_{\rm{tail}}$ fitting algorithm, particularly Eq. \ref{eq:density-profile-SI}, is a good generalization of previous density profile approximations of FDM halos.

One interesting observation from the data presented in Figure \ref{fig:high-res-evolution} is the slow, secular evolution of the density profile at both small and large radii. Regardless of SI strength, the soliton central density and the tail power-law index increase slowly from $t_{\rm{merge}}=0.7~\rm{Gyr}$ through the simulation end. These ``compactifying'' and ``shallowing'' trends indicate that dark matter is redistributing itself within the halo, and in particular, mass is increasing in the center and decreasing in the outer envelope. Presumably, the trends would continue if the simulation were extended, but level off at some equilibrium state later on.

With the suite of four analogous simulations at an intermediate resolution $N=200$ evolved to a much later $T = 20~\rm{Gyr}$, we can gather hints about the longer-term evolution of these halos. Figure \ref{fig:long-term-evolution} shows the results of these mid-resolution simulations in the same format as the top panels of Figure \ref{fig:high-res-evolution}. The simple moving averages are highlighted to isolate secular trends and reduce noise, and 4 Gyr is marked with a vertical line to indicate the end of the higher-resolution runs. Notably, a collapse is observed in the $f_{15}=1.2$ simulation at $t = 5.87~\rm{Gyr}$; this is discussed further in Section \ref{sec:strongly-self-interacting-halos}. For the others, the general increase of $\rho_0$ and $n$ is observed to extend far past $t = 4~\rm{Gyr}$. In the $f_{15}=\infty$ and $1.5$ simulations, the data suggests that an equilibrium state is not reached since either $\rho_0$ or $n$ is steadily changing throughout. By contrast, neither $\rho_0$ nor $n$ increase significantly after $t \sim 5~\rm{Gyr}$ in the $f_{15}=2.0$ simulation, so that halo may have reached equilibrium. At the end of these extended simulations, the trends from Figure \ref{fig:high-res-evolution} are still apparent: the central density depends on interaction strength while the power-law index of the outer envelope does not.

\subsubsection{Trends with Self-Interaction Strength}
\label{sec:trends-with-SI-strength}

\begin{figure*}
    \centering
    \includegraphics[width=\textwidth]{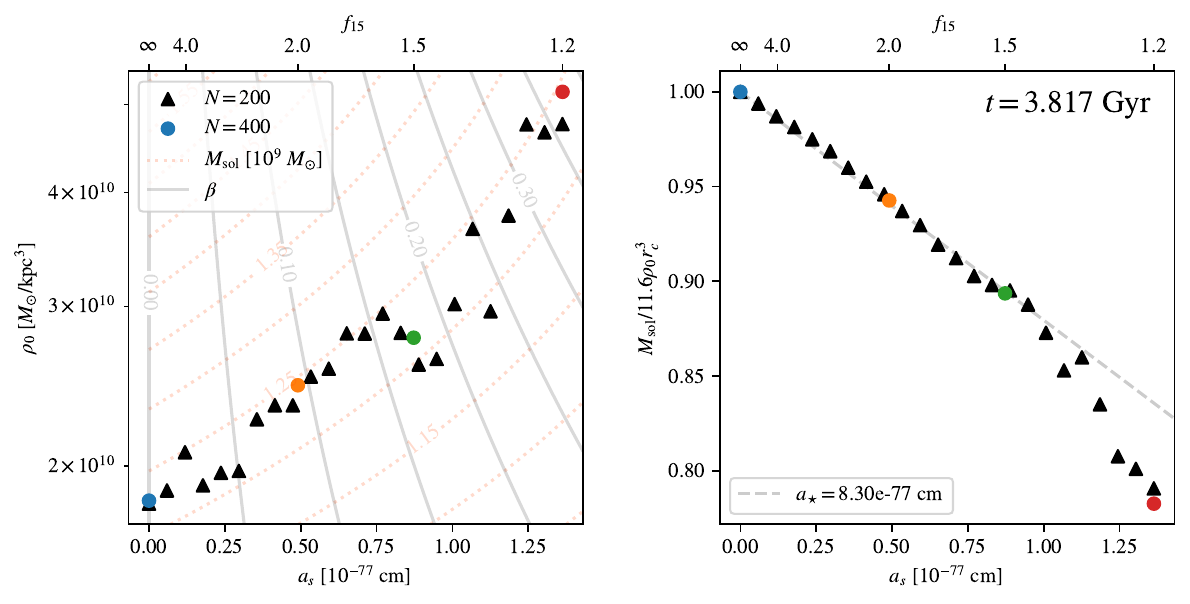}
    \caption{Variation of soliton properties across self-interaction strength using both our 4 higher-resolution simulations ($N=400$, same colors as previous figures) and our 24 lower-resolution runs ($N=200$, black triangles). \textit{Left}: The soliton central density increases with stronger self-interactions, as suggested by the snapshots in e.g. Figure \ref{fig:density-profile-comparison}. The trend steepens as the interaction strength nears the critical point at which the soliton collapses. At $f_{15}=1.2$, $\rho_0$ is approximately 3 times higher than in the non-interacting case. The solid gray contours show constant values of $\beta$ while the red dashed contours show constant values of $M_{\rm sol}$. The contours demonstrate that $\beta$ increases quickly with SI strength; by contrast, $M_{\rm{sol}}$ stays approximately constant. \textit{Right}: Soliton mass, normalized by the prediction from Eq. \ref{eq:soliton-mass}. The residual dependence on $a_s$ is roughly linear up until $f_{15} \approx 1.5$, at which point the decrease steepens.}
    \label{fig:correlations-with-a_s}
\end{figure*}

Figures \ref{fig:density-profile-comparison}, \ref{fig:high-res-evolution} and \ref{fig:long-term-evolution} suggest that weakly attractive self-interactions smoothly change soliton properties. To see these changes more clearly, we extract and fit density profiles from our suite of 24 intermediate-resolution simulations spanning the weak SI regime. We continue to hold the initial configuration constant among each simulation; the only difference is the strength of self-interactions.

To be clear, the way in which the soliton density profile changes with $\rho_0$ and $a_s$ is already accurately quantified in Eq. \ref{eq:density-profile-SI}. What we hope to glean from these simulations is how $\rho_0$ changes with $a_s$ for a fixed initial mass configuration. For our initial conditions, $a_s$ parameterizes a particular curve $\rho_0(a_s)$ that traces a slice of $M_{\rm{sol}}$ or $\beta$ heatmaps. In this sense, $M_{\rm{sol}}$ and $\beta$ may be considered solely functions of $a_s$ for our simulation setup, since $\rho_0$ is determined at every $a_s$ (i.e., $M_{\rm{sol}}(a_s) \equiv M_{\rm{sol}}(\rho_0(a_s), a_s)$). However, if our simulation had different initial conditions, the relationship $\rho_0(a_s)$ would be different and thus $M_{\rm{sol}}(a_s)$ and $\beta(a_s)$ would look different.

The left panel of Figure \ref{fig:correlations-with-a_s} plots $\rho_0$, the one free parameter in the fit to the soliton density profile, as measured at the end of each of these 24 simulations, as well as the 4 higher-resolution simulations introduced in previous subsections. The central density is sampled in each simulation from a density profile averaged over three consecutive snapshots to mitigate random fluctuations. The measurements confirm the trends suggested in previous figures: as compared to the collisionless case, $\rho_0$ increases smoothly with stronger self-interactions (and the core radius decreases according to Eq. \ref{eq:rho0-r_c-relation}). The trend is not linear in $\rho_0$ or $\log\rho_0$ but more closely resembles an asymptotic or exponential increase. Near the critical SI strength at which our soliton collapses ($f_{15}\sim1.1$ to $1.2$), the central density is approximately 3 times greater than in the non-interacting case.

Overlaid are contours of equal $\beta$ and contours of equal $\log M_{\rm{sol}}$. As interaction strength increases, $\beta$ increases quickly since both multiplicative terms in Eq. \ref{eq:beta} are increasing. The changes to $M_{\rm{sol}}$ are a bit more complicated. From numerical integrals of Eq. \ref{eq:density-profile-SI}, we know that increasing $a_s$ generally decreases the soliton mass (see shape changes in the right panel of Figure \ref{fig:density-profile-comparison}), while increasing $\rho_0$ increases $M_{\rm{sol}}$ (typical $M_{\rm{sol}} \propto \rho_0^{1/4}$ growth in FDM halos). In our simulations, $\rho_0$ increases in response to an increase in $a_s$, so the resulting soliton mass is a product of two competing effects. We find that self-interactions increase $\rho_0$ in such a way that $M_{\rm{sol}}$ is nearly constant over our range of simulations, increasing only slightly from $1.20$ to $1.25 \times 10^{9}~M_{\odot}$.

To isolate and quantify the soliton mass changes that differ from the typical growth predicted by Eq. \ref{eq:soliton-mass}, we can normalize the measured $M_{\rm{sol}}$ at each $a_s$ by the zero-SI prediction $M_{\rm{sol}}(a_s=0) = 11.6\rho_0r_c^3$. The right panel of Figure \ref{fig:correlations-with-a_s} unveils a tight residual linear dependence of $M_{\rm{sol}}$ on $a_s$ up to $f_{15}=1.5$. This suggests that for very weak self-interactions in our simulations, the soliton mass can be accurately predicted by revising Eq. \ref{eq:soliton-mass} to
\begin{equation} \label{eq:soliton-mass-SI-2}
    M_{\rm{sol}}(\rho_0,a_s) = 11.6\rho_0 r_c^3\left( 1 - \frac{a_s}{a_{\star}}\right)
\end{equation}
with one additional constant $a_{\star}$ characterizing the strength of the dependence. With our particular initial conditions, we find $a_{\star}\approx 8.30\times 10^{-77}~\rm{cm}$ (which corresponds to $f_{15,\star}\approx0.49$); the fit line with that slope is also plotted on the right panel. Note that if Eq. \ref{eq:soliton-mass-SI-2} is true and $M_{\rm{sol}}$ is observed to remain roughly constant over a range of $a_s$, then $\rho_0 \propto (1-a_s/a_{\star})^{-4}$. After $f_{15}=1.5$, the $a_s$ dependence departs from linear. Note that even though $a_{\star}$ is formulated to look like a characteristic SI strength, it lies well outside the domain of weak interactions. It is always the case that $a_s \ll a_{\star}$, so self-interactions can only fractionally reduce the soliton mass.

\subsubsection{Energy Profiles}
\label{sec:energy-profiles}

\begin{figure*}
    \centering
    \includegraphics[width=\textwidth]{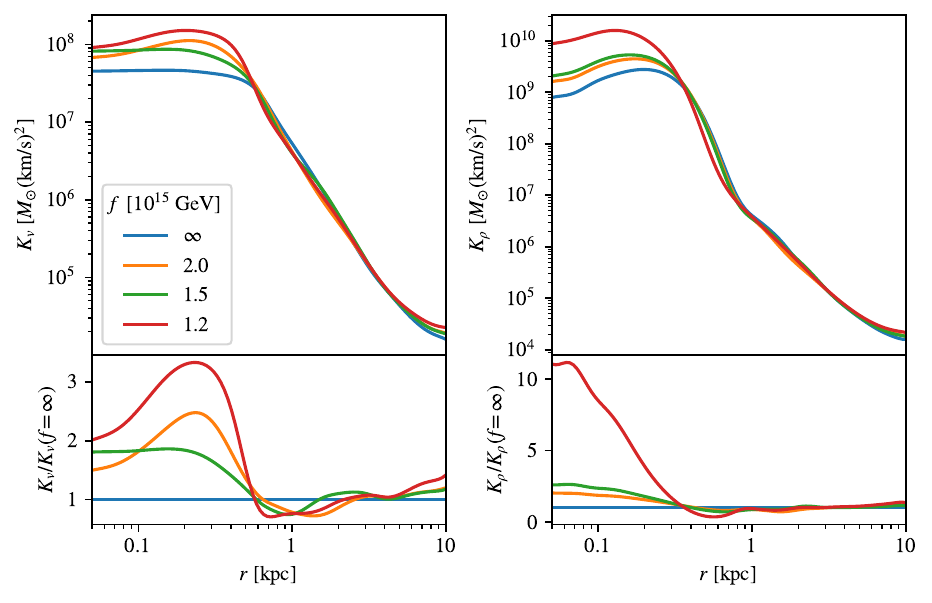}
    \caption{Classical kinetic energy $K_v$ (\textit{left}) and quantum gradient energy $K_{\rho}$ (\textit{right}) profiles at the end of each $N=400$ simulation with weak self-interactions. Similar to the trends seen in the density profiles, stronger self-interactions enhance both components of the kinetic energy around the soliton with some degree of dependence on the interaction strength. In the outer envelope, the slope and amplitude of kinetic energy profiles are independent of self-interactions.}
    \label{fig:energy-profiles}
\end{figure*}

As with the density, the quantum gradient energy $K_{\rho}$, and classical kinetic energy $K_v$ can be computed at each grid cell within the simulation box and may be decomposed into radial profiles. This analysis was considered in \cite{mocz2017} and is useful for determining the energy composition of different regions of the halo and offers evidence as to the forces at play and the mechanisms driving the halo evolution.

Figure \ref{fig:energy-profiles} shows kinetic energy profiles at the end of our weakly self-interacting $N=400$ simulations. The main features of the profiles are preserved regardless of $f_{15}$: mass and energy is most concentrated in the soliton with curves characteristic of FDM, and the curves fall off with power laws after a certain cutoff radius. However, the amplitude of each quantity at low radii is dependent on self-interactions. Stronger self-interactions enhance both kinetic energy components within the soliton core radius. In the left panel, the $K_v$ increases interior to $r \sim 0.5~\rm{kpc}$ when self-interactions are present, and the increase is sorted by SI strength interior to $r \sim 0.1~\rm{kpc}$ as in the density profiles. Similarly, in the right panel, $K_{\rho}$ is measured to increase interior to $r \sim 0.3~\rm{kpc}$ to some extent dependent on SI strength. In the $f_{15}=1.2$ simulation, $K_{\rho}$ is enhanced by more than an order of magnitude in the soliton center. In both cases, the energy profile of the outer envelopes is invariant with interaction strength, further indicating that self-interactions are not prominent at these radii.

Both kinetic energy components maximize at $r \sim r_c$ and decrease to a local minimum at the soliton center, regardless of interaction strength.

\subsection{Strongly Self-Interacting Halos}
\label{sec:strongly-self-interacting-halos}

The analysis so far has included only dilute solitons -- solitons affected by particle self-interactions but not to the point of collapse predicted in Eq. \ref{eq:critical-soliton-mass}. However, we observe soliton collapses in three simulations with strong enough self-interactions. Two of these simulations are part of our higher-resolution $N=400$ suite ($f_{15}=1.1$ and $1.0$) while the other is one of our extended lower-resolution runs ($f_{15}=1.2$). In all of these simulations, the transition happens quickly ($\Delta t < 10$ Myr) and the resulting ``compact'' soliton is spatially unresolved ($r < 50$ pc), appearing as one single dense pixel. After the collapse event, numerical artifacts of poor resolution are propagated throughout the box, reducing the credibility of the data on the outer envelope. However, we can analyze the simulations before the soliton collapses, as well as perform tests by modifying the physics to artificially lower $\rho_{\rm{eq}}$ to a resolvable level.

\subsubsection{Phase Transition}
\label{sec:phase-transition}

\begin{figure*}
    \centering
    \includegraphics[width=\textwidth]{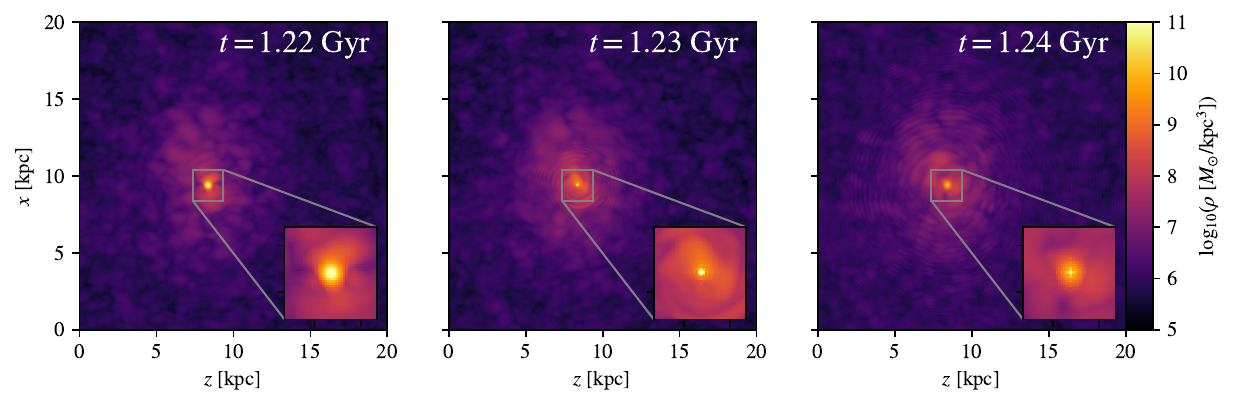}
    \caption{Soliton collapse under strong axion self-interactions ($f_{15}=1.1$). \textit{Left}: Just before collapsing, the soliton and outer envelope are typical of other snapshots of dilute solitons with no obvious indicators of an imminent collapse. \textit{Center}: Immediately after collapse, the soliton appears as a dense, tiny unresolved object with waves emanating from its center. \textit{Right}: Unphysical numerical artifacts from the unresolved soliton and collapse wavefronts blur and distort the outer envelope, rendering any post-collapse data unreliable.}
    \label{fig:collapse}
\end{figure*}

Soliton collapse is readily apparent in density projections of the $f_{15}=1.1$ and $1.0$ simulations over a few select snapshots. Figure \ref{fig:collapse} shows a sequence of three consecutive projections from the $f_{15}=1.1$ simulation at our highest temporal resolution of $\Delta t = 10$ Myr. In the left panel at $t=1.22$ Gyr, the soliton is dilute, as it has been since the initial configuration fully merged together at $t \approx 0.7$ Gyr. There are no unusual features in the halo that indicate an imminent collapse, other than the fact that the soliton mass ($M_{\rm{sol}} = 1.22 \times 10^9~M_{\odot}$) and central density ($\rho_0 = 1.85 \times 10^{11}~M_{\odot}/\rm{kpc}^3$) are at their highest values yet, well within the critical regime for $f_{15}=1.1$. In the next $10$ Myr (between the left and center snapshots), the soliton fully transitions to its compact state, displayed as a cross of five dense pixels in the center panel. Spherical waves are observed to emanate from the center at a velocity of $v \sim 10^3~\rm{km}/\rm{s}$ with various frequencies, although it is unclear whether these are physical or not. By $t=1.24$ Gyr in the right panel, the compact soliton appears unphysically as one dense pixel surrounded by a small overdense cloud. The ripples have propagated through the periodic boundaries, distorting the granules and blurring vortex lines within the box. During and after the collapse, neither the slope nor amplitude of the power-law outer envelope is changed. Further, it extends to fill the void left by the dilute soliton, leaving a cuspier halo profile. However, we stress that these results are not rigorous; simulations that adequately resolve the soliton collapse are required to thoroughly investigate the post-collapse halo structure.

We can justify the observation that the compact soliton is smaller than our grid resolution by noting other length scales relevant to the system. Equation \ref{eq:equilibrium-density} suggests that the post-collapse equilibrium density of the compact soliton should be $\rho_{\rm{eq}} \sim 3.7 \times 10^{17}\, M_{\odot}/\rm{kpc}^3$, over six orders of magnitude higher than $\rho_0$, pre-collapse. Extrapolating the $\rho_0$-$r_c$ relation in Eq. \ref{eq:rho0-r_c-relation} to the post-collapse regime yields a compact radius of $r = 2.7$ pc, about $20$ times finer than our best grid resolution. However, there is no evidence to suggest that Eq. \ref{eq:rho0-r_c-relation} can be extrapolated in this way. If the mass of the compact soliton equals the mass of its dilute progenitor, and if the compact soliton can be approximated to a sphere of uniform density $\rho_{\rm{eq}}$, then the post-collapse radius is $r \sim (3M_{\rm{sol}}/4\pi\rho_{\rm{eq}})^{1/3} = 0.9$ pc, another factor of three finer resolution. These lengths are about $4$ orders of magnitude higher than the Schwarzschild radius of the soliton $R_{\bullet} = 2GM_{\rm{sol}}/c^2 = 1.1 \times 10^{-4}$ pc, indicating that the higher-order $|\psi|^6$ repulsive pressure may halt the collapse before a black hole is formed.

Similarly, the free-fall time of the dilute soliton may give an order-of-magnitude estimate of the collapse timescale. Using the pre-collapse central density, the free-fall time is $t_{\rm{ff}} \sim (G\rho_0)^{-1/2} = 1.2$ Myr, about $10$ times finer temporal resolution than our smallest $\Delta t$ between timesteps.

Figure \ref{fig:collapse-thresholds} plots the dimensionless self-interaction parameter $\beta$, which is proportional to $\rho_0^{1/2}$, over time for each of the $f_{15}=1.2,1.1$, and $1.0$ simulations up to the point of collapse. The gray shaded region indicates the period during which merging events take place and the $\times$ markings and vertical dotted lines indicate the moments right before each phase transition occurs. In this limited set of data, solitons collapsed first in simulations with stronger self-interactions. The $f_{15}=1.0$ soliton collapsed during the last major merging event, before the dilute soliton had a chance to condense fully. The $f_{15}=1.1$ soliton was dilute for a short period of $\sim 0.5$ Gyr after merging finished. The $f_{15}=1.2$ soliton was dilute for much longer, slowly accreting mass for $\sim 5$ Gyr before finally collapsing.

Notably, finer temporal sampling reveals that the central density in the $f_{15}=1.1$ simulation increases immediately before collapse at an unusually fast pace. Whereas $\rho_0$ evolves slowly in other simulations with dilute solitons, in this case it violently oscillates up by a factor of $\sim 4$ over $< 200$ Myr from $t=1.0$ to $1.2$ Gyr. This stage of evolution is distinct from collapse, which happens much more quickly; the central density briefly accelerates while the soliton is still dilute before reaching some threshold for collapse. We did not perform finer sampling of the $f_{15}=1.2$ simulation immediately before collapse; the timesteps remain at a default $\Delta t = 100$ Myr. A similarly steep ascent in density could be revealed with more detail, though this simulation was performed at a lower resolution, so it may not be resolved as accurately. 

This data indicates only a very tight range of interaction strengths ($1.0 \lesssim f_{15} \lesssim 1.5$, for our halo mass) allows for a dilute soliton to form before collapsing at some point later in its lifetime. If $f_{15}$ is higher, the soliton will remain dilute forever, but if $f_{15}$ is lower, a collapse will occur before the soliton fully forms in a dilute state. Put differently, for a boson mass of $m \sim 10^{-22}$ eV and interaction strength near $f \sim 10^{15}$ GeV, the Universe would likely contain both dilute and compact solitons with collapses actively occurring as solitons accrete mass and surpass their critical threshold. If self-interactions are weaker, there may be no compact solitons in the Universe, but if they are stronger, \textit{every} halo may contain a compact remnant.

\subsubsection{Threshold Criteria for Collapse}
\label{sec:collapse-criteria}

\begin{figure}
    \centering
    \includegraphics[width=\columnwidth]{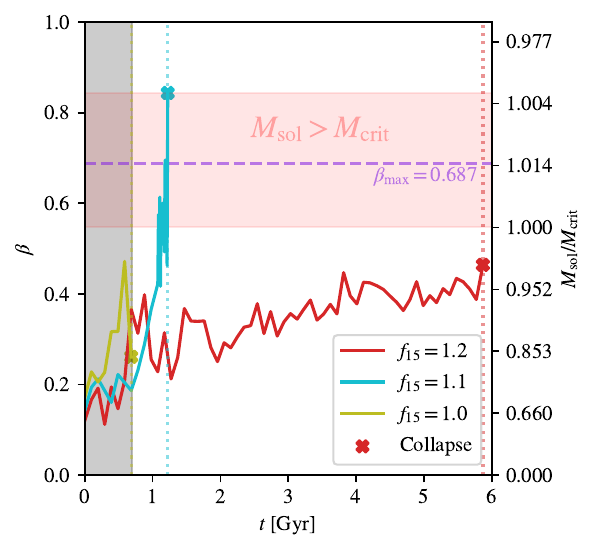}
    \caption{Stability parameter $\beta$ tracked over time for each of the three simulations in which we observed a soliton collapse. In the red shaded region, $M_{\rm{sol}} > M_{\rm{crit}}$; the maximum $M_{\rm{sol}}/M_{\rm{crit}}$ ratio is marked at $\beta_{\rm{max}}=0.687$, and the time period before all halos fully merge together is shaded gray. The $f_{15}=1.2$ soliton collapses at its highest-yet value of $\beta=0.464$, where $M_{\rm{sol}} \approx 0.97M_{\rm{crit}}$, after a gradual $\sim 5$ Gyr increase. By contrast, the $f_{15}=1.1$ soliton collapses right on the upper boundary of the $M_{\rm{sol}} > M_{\rm{crit}}$ region, at $\beta=0.844$, after a steep $\sim 0.5$ Gyr ascent. The $f_{15}=1.0$ simulation collapses during the final major subhalo collision; finer timesteps would be needed to resolve the details.}
    \label{fig:collapse-thresholds}
\end{figure}

The critical soliton mass, given by Eq. \ref{eq:critical-soliton-mass}, offers a predictive threshold for soliton collapse, which we extended to other criteria in Eqs. \ref{eq:beta-maximizer}, \ref{eq:rho-crit}, and \ref{eq:r_c-crit} from numerical analysis of the ground state solution of the GPP equations. It is of interest to examine whether these criteria accurately predicted collapse in our simulations.

Figure \ref{fig:collapse-thresholds} includes corresponding $M_{\rm{sol}}/M_{\rm{crit}}$ ratios on the right-hand axis, according to Eq. \ref{eq:soliton-mass-ratio-beta-correspondence}. The red-shaded region indicates the critical regime ($0.55 < \beta < 0.84$) where $M_{\rm{sol}} > M_{\rm{crit}}$, and the maximum predicted soliton mass is marked with a dashed purple line. The $f_{15}=1.1$ and $1.2$ simulations collapse on opposite sides of the critical regime, at quite different values of $\beta$. The $f_{15}=1.2$ soliton collapses after rising from $\beta \sim 0.3$ to $\beta = 0.46$ over $5$ Gyr, reaching $M_{\rm{sol}} \approx 0.97M_{\rm{crit}}$ but never quite surpassing the critical mass. We stress, however, that lower resolution in this simulation may instigate soliton collapse prematurely. On the other hand, the $f_{15}=1.1$ soliton ascends quickly through the critical regime to $\beta=0.84$. Interestingly, the quick ascent triggers close to the lower $M_{\rm{sol}}=M_{\rm{crit}}$ boundary, and the collapse occurs right on the upper boundary. We conclude that the criteria given in Eq. \ref{eq:critical-soliton-mass} accurately predicts soliton collapse in our simulations to within a few percent.

\subsubsection{Boosting the $|\psi|^6$ Repulsive Pressure}
\label{sec:boosting-the-repulsive-term}

\begin{figure*}
    \centering
    \includegraphics[width=\textwidth]{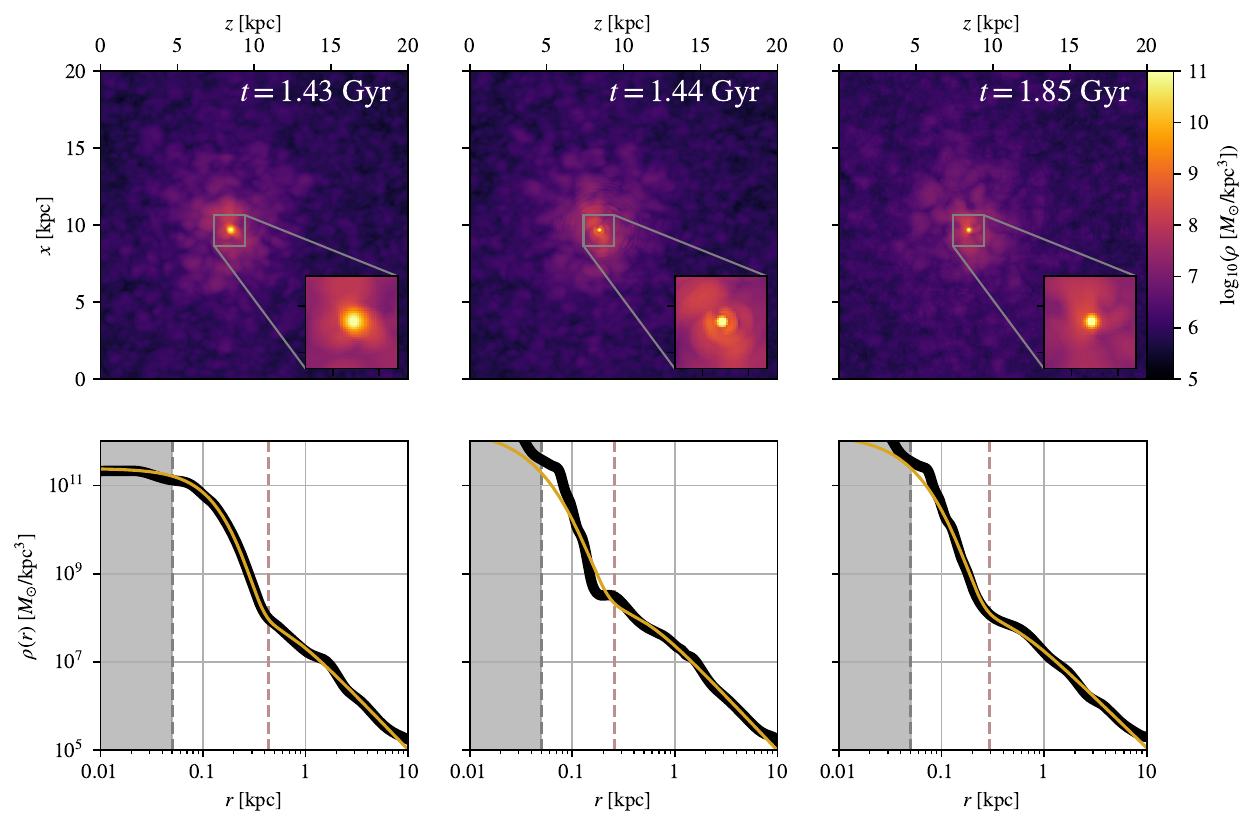}
    \caption{\textit{Top}: Soliton phase transition under artificially modified physics that lowers the post-collapse equilibrium density by five orders of magnitude. In this modified regime, the compact soliton is resolved to have a diameter of $\sim 0.3~\rm{kpc}$, or about $6$ pixels and persists in a stable state through the end of the simulation. \textit{Bottom}: The density profile of the compact soliton (thick black curves, center and right) in this modified physics simulation is not as well-fit by the same fitting function (Eq.~\ref{eq:density-profile-SI}) used for dilute solitons; the slope is too steep. However, the information is still very limited and it is unclear how these results extrapolate to a fully collapsed object (with unmodified physics).}
    \label{fig:collapse-with-repulsive-boost}
\end{figure*}

In an attempt to resolve the compact soliton with our highest spatial resolution of $\Delta x = 0.05$ kpc, we artificially increase the higher-order stabilizing pressure term in Eq. \ref{eq:gross-pitaesvkii-poisson-corrected} and rerun the $f_{15}=1.1$ simulation with this change. Multiplying the higher order term by a ``boosting'' factor $B \gg 1$ increases the repulsive pressure and halts collapse at a lower equilibrium density
\begin{equation}
\label{eq:equilibrium-density-modified}
    \rho_{\mathrm{eq, modified}} = \frac{\rho_{\mathrm{eq}}}{B}
\end{equation}
and thus a larger compact soliton radius. In the $f_{15} = 1.1$ simulation, the dilute soliton is observed to have a central density of $\rho_0 \approx 1.85 \times 10^{11} M_{\odot}/\rm{kpc}^3$, six orders of magnitude less than $\rho_{\rm{eq}} \approx 3.7 \times 10^{17} M_{\odot}/\rm{kpc}^3$. We find a boosting factor of $B = 10^5$ to be large enough to resolve the compact object while small enough that a collapse still initiates.

For direct comparison, we initialize a new simulation with the $t = 1.08$ Gyr snapshot from the default-physics $f_{15}=1.1$ simulation. We evolve forward for $0.8$ Gyr at a finer temporal resolution of $\Delta t = 10$ Myr.

Figure \ref{fig:collapse-with-repulsive-boost} shows a sequence of density projections analogous to Figure \ref{fig:collapse}, but with the repulsive pressure boosted. The soliton remains dilute up through the $t=1.43$ Gyr snapshot (left panel; $0.21$ Gyr later than with default physics) at which point the central density is $\rho_0 = 2.01 \times 10^{11}~M_{\odot}/\rm{kpc}^3$, the mass is $M_{\rm{sol}}=1.22 \times 10^{9}~M_{\odot}$ and $\beta=0.88$. Relative to the default $f_{15}=1.1$ simulation, $\rho_0$ is slightly higher, presumably because the boosted higher-order pressure offers a slight additional resistance against collapse. The phase transition occurs quickly between $t=1.43$ and $1.44$ Gyr, leaving behind a compact soliton half the size of its dilute parent with an order of magnitude higher central density, $\rho_0 \approx 2.8 \times 10^{12}~M_{\odot}/\rm{kpc}^3$. This is within a factor of two of the equilibrium density, $\rho_{\rm{eq,modified}}=3.7 \times 10^{12}~M_{\odot}/\rm{kpc}^3$ as calculated by Eq. \ref{eq:equilibrium-density-modified}. Shockwaves similar to those observed in Figure \ref{fig:collapse} are also observed emanating from the compact object. The soliton remains stably in its compact state for the remainder of the simulation. Density profiles are shown in the lower panels of Figure \ref{fig:collapse-with-repulsive-boost}; the dilute soliton fitting formula, Eq. \ref{eq:density-profile-SI}, struggles to fit the steep slope of the compact soliton at $r=0.08$--$0.2$ kpc. However, even in this modified regime, the information is very limited: the compact soliton is only resolved by $\sim 7$ pixels in diameter. Furthermore, it is unclear how these results extrapolate to a fully collapsed object with unmodified physics. 


\section{Discussion}
\label{sec:discussion}

The simple one-parameter FDM model has likely been excluded as a dominant component of dark matter in the Universe because it is unable to reconcile extended cores in dwarf galaxy dark matter halos with adequate structure observed in the Lyman-$\alpha$ forest. However, more general models of wave dark matter may offer the same benefits of FDM (e.g., fewer satellite galaxies, cored halos, natural production in the early universe with the correct abundance) while boosting small-scale power. Attractive axion self-interactions are one such extension, offering a critical mass scale above which the inner part of the dark matter halo contracts into a small, dense object.

Depending on their strength, attractive self-interactions can imprint two distinct signatures on an FDM halo. If the self-interactions are weak, the soliton density profile will change very slightly, becoming more compact but with a slightly less extended core than predicted in the simple FDM model. Currently, rotation curves of nearby galaxies are not measured precisely enough to constrain the self-interaction parameter by discerning between density profiles; the change is too small and only present in the inner parsec of the halo (the outer NFW-like envelope is not affected by self-interactions). However, if the self-interactions are strong, every soliton in the Universe above a certain mass is expected to collapse into either a boson star supported by the higher order $\psi^6$ repulsive term or a black hole; high-resolution fully general relativistic simulations are required to determine the outcome. Further assuming a positive halo mass--soliton mass correlation, every dark matter halo above a certain mass would be expected to contain one of these objects---either a very dense dark matter environment at its center or a supermassive black hole (SMBH)---seeded by soliton collapse \citep{padilla2021}. Correlations between observed SMBHs in galaxies and their host dark matter halos may be used to constrain these theories. The collapse itself may be accompanied by a ``bosenova'', a huge flux of relativistic axions that may form a cosmic axion background, similar to a gravitational wave background from merging SMBHs, if such collapses are frequent enough. More careful simulations of soliton collapse, along with the post-collapse interplay between the dense remnant and the dark matter halo, are very important to establish the full predictions of the model.

In our simulations, we can characterize the effect of self-interactions on dilute solitons, but we are not able to resolve soliton collapse without modifying the physics. However, we do observe the halo in snapshots before and during collapse. Based on pre-collapse soliton mass values, we find a rough agreement (within $\sim 3\%$) with the critical mass formula, Eq. \ref{eq:critical-soliton-mass}, developed in \cite{chavanis2011, chavanis2011a}. Equation \ref{eq:critical-soliton-mass} assumes a static, isolated initial condition and should be considered an approximation for our simulations. During collapse, we observe ripples emanating from the soliton that propagate throughout the box and distort the granule fluctuations. While it is not clear whether this is simply an artifact of poor resolution, it is clear that a new, smaller scale is introduced into the halo. If the ripples are physical, it may be necessary to simulate the envelope, in addition to the core, in high resolution to capture the post-collapse equilibrium configuration.

Besides spatial and temporal resolution, our simulations are limited in that we test only one initial mass configuration. It would be interesting to confirm that higher or lower mass solitons still obey the generalized density profile, Eq. \ref{eq:density-profile-SI}, and the critical mass formula, Eq. \ref{eq:critical-soliton-mass}.  


\section{Conclusion}
\label{sec:conclusion}

In this work, we investigate the consequences of including attractive self-interactions in the Gross-Pitaevskii-Poisson equations for ultralight axion dark matter by simulating idealized dark matter halos under various SI strengths. We summarize our principal findings as follows:
\begin{itemize}
    \item Attractive self-interactions introduce two distinct soliton phases: a ``dilute'' phase when self-interactions are weak and a ``compact'' phase when self-interactions are strong. A soliton can transition from a dilute to compact state by accreting matter, exceeding a critical mass threshold, and collapsing. 
\end{itemize}

\noindent In the weak self-interaction regime:
\begin{itemize}
    \item Solitons formed from identical initial conditions change shape to some degree dependent on the particle interaction strength. Stronger self-interactions \textit{increase} the central density $\rho_0$ and \textit{decrease} the core radius $r_c$ relative to the collisionless case. The observed decrease in $r_c$ very nearly follows the expected $r_c \propto \rho_0^{-1/4}$ dependence in Eq. \ref{eq:rho0-r_c-relation}, but a slight correction is observed to be non-negligible (see Figure \ref{fig:density-profile-comparison}).
    \item We find that the soliton density profile is accurately generalized by the fitting formula in Eq. \ref{eq:density-profile-SI}. This approximation is calibrated on numerical analysis of the ground state solution of the GPP equations and validated by our simulations.
    \item The soliton mass $M_{\rm{sol}}$ is nearly constant as self-interactions are dialed up. Typical $M_{\rm{sol}} \propto \rho_0^{1/4}$ dependence is nearly counterbalanced by deviations in the density profile seen in the right panel of Figure \ref{fig:density-profile-comparison}. For very weak self-interactions ($f_{15} \geq 1.5$), $M_{\rm{sol}}$ can be approximated by Eq. \ref{eq:soliton-mass-SI-2}, a generalization of Eq. \ref{eq:soliton-mass} with a simple linear dependence on the $s$-scattering length.
    \item The outer envelope of the halo that surrounds the soliton is invariant of SI strength. Evidence suggests this lack of dependence extends to the strong interaction regime, as neither the slope nor amplitude of the outer envelope density profile appears to change after a soliton phase transition.
\end{itemize}

\noindent In the strong self-interaction regime:
\begin{itemize}
    \item The time period during which a soliton is dilute before collapsing depends on interaction strength. If interactions are too weak, the soliton will never collapse. If interactions are too strong, as in our $f = 1.0 \times 10^{15}$ GeV run, the dense center will collapse before a dilute soliton has a chance to form. In a very tight range, for us $1.0 \times 10^{15} \lesssim f \lesssim 1.2 \times 10^{15}~\mathrm{GeV}$, a dilute soliton can form before later transitioning to its compact state.
    \item Immediately before collapsing, a dilute soliton may undergo a runaway ascent in central density unlike the secular evolution observed when $M_{\rm{sol}}$ is significantly less than $M_{\rm{crit}}$ (see $f = 1.1 \times 10^{15}$ GeV simulation in Figure \ref{fig:collapse-thresholds}).
    \item The collapse happens very quickly relative to our simulation length and snapshot spacing, with upper limit $\Delta t_{\rm{collapse}} < 10$ Myr. The compact remnant is also unresolved in our simulations, but additional tests with modified physics indicate that the compact soliton will have central density close to the equilibrium density, given in Eq. \ref{eq:equilibrium-density}.
\end{itemize}

Naturally predicted axion-like self-interactions with a symmetry-breaking scale $f \sim 10^{15}$ GeV can significantly alter the predictions made from the simple FDM model. For strong enough self-interactions, nearly every DM halo in the Universe would be expected to contain compact soliton, possibly in the form of a supermassive black hole, which could spawn strong non-linear effects on small scales and remedy the \textit{Catch-22} that faces FDM. As the SI strength is dialed down, higher-mass halos are expected to retain dilute solitons, which smoothly approximate vanilla FDM solitons as $f \rightarrow \infty$. From here, the most important next step is to carefully simulate soliton collapse with full general relativistic effects to develop a seeding formula for cosmological simulations. With models for soliton phase transition and post-collapse interactions with the host halos, large-scale simulations can determine whether attractive self-interactions broaden FDM parameter space enough satisfy modern observations of both dwarf galaxies rotation curves and small-scale structure in the Lyman-$\alpha$ forest. 


\section*{Acknowledgments}

CAP was partially supported by NSF grant AST-2108962. MBK acknowledges support from NSF CAREER award AST-1752913, NSF grants AST-1910346 and AST-2108962, NASA grant 80NSSC22K0827, and HST-AR-15809, HST-GO-15658, HST-GO-15901, HST-GO-15902, HST-AR-16159, HST-GO-16226, HST-GO-16686, HST-AR-17028, and HST-AR-17043 from the Space Telescope Science Institute, which is operated by AURA, Inc., under NASA contract NAS5-26555. P.M. acknowledges this work was in part performed under the auspices of the U.S. Department of Energy by Lawrence Livermore National Laboratory under contract DE-AC52-07NA27344, Lawrence Livermore National Security, LLC.

\appendix

\section{Accuracy of Generalized Density Profile Fitting Formula}
\label{sec:accuracy-appendix}

\begin{figure}
    \centering
    \includegraphics[width=\columnwidth]{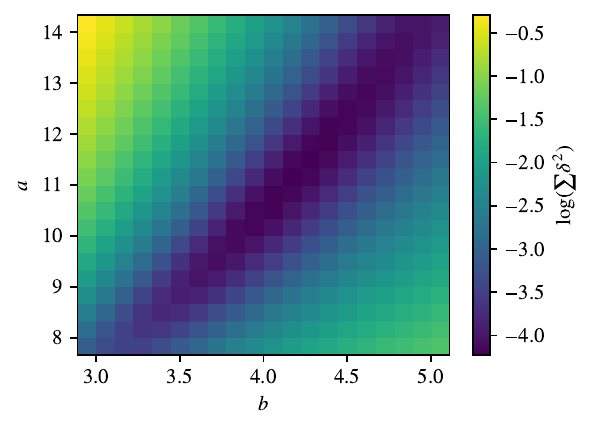}
    \caption{Goodness of fit measurement for Eq. \ref{eq:density-profile-SI} using many values of parameters $a$ and $b$. The optimal combination is $a=11.2$ and $b=4.2$, and these values are set constants throughout the entire analysis.}
    \label{fig:residuals-grid}
\end{figure}

\begin{figure}
    \centering
    \includegraphics[width=\columnwidth]{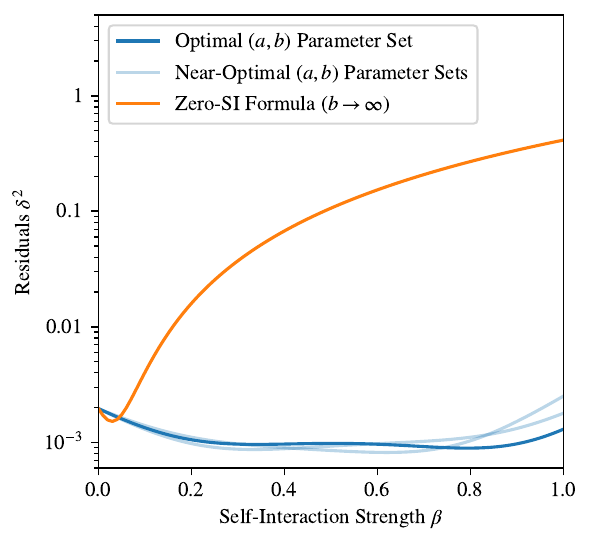}
    \caption{Comparison of goodness of fit of the zero-SI formula, Eq. \ref{eq:density-profile-no-SI} (orange), to the generalized formula, Eq. \ref{eq:density-profile-SI} (blue), over the relevant range of SI strengths. The soliton density profile changes subtly with self-interactions such that Eq. \ref{eq:density-profile-SI} yields a much more accurate description. The optimal $(a,b)$ parameter set $(11.2,4.2)$ is shown as the bolded curve, but other $(a,b)$ parameter sets lying along the degeneracy (see Figure \ref{fig:residuals-grid}) are also included: $(10.5,3.9)$ and $(12.5,4.5)$.}
    \label{fig:residuals-vs-beta}
\end{figure}

In Eq. \ref{eq:density-profile-SI}, we propose a new fitting formula for dilute solitons under attractive self-interactions that generalizes Eq. \ref{eq:density-profile-no-SI} originally found in \cite{schive2014a}. First, we seek to identify the $(a,b)$ parameter set that best matches solutions to the GPP equations over the relevant range of interaction strengths. We do this via an optimization process, visualized in Figure \ref{fig:residuals-grid}. For any two values of $a$ and $b$, the goodness of fit metric $\delta^2$ from Eq. \ref{eq:goodness-of-fit} can be used to compare our predictions to the GPP solutions across a range of $\beta$ values. The total accuracy of an $(a,b)$ model is simply the sum of $\delta^2$ values over $0 < \beta < 1$. In Figure \ref{fig:residuals-grid}, these $\sum\delta^2$ accuracy values are displayed by color for a $20 \times 20$ grid of $(a,b)$ values. The optimal parameter set is found to be $a=11.2$ and $b=4.2$, but there is a strong degeneracy between the two parameters ($a \approx 4b-5.5$). Increasing $a$ tends to decrease the density over the relevant range of radii, while increasing $b$ reverses the effect. In this sense, the functional form of Eq. \ref{eq:density-profile-SI} admits a family of well-fitting solutions to self-interacting FDM halos.

To demonstrate the accuracy of our fitting formula with its optimal $(a,b)$ parameter set, Figure \ref{fig:residuals-vs-beta} plots the residuals $\delta^2$ as a function of $\beta$. Also included are residual functions for the zero-SI formula, Eq. \ref{eq:density-profile-no-SI}, and two nearly-optimal $(a,b)$ parameter sets that lie along the degeneracy in Figure \ref{fig:residuals-grid}. While the fit using the zero-SI formula becomes increasingly poor as the SI strength is dialed up, Eq. \ref{eq:density-profile-SI} offers persistently great fits to the GPP solutions. The two other $(a,b)$ parameter sets shown, $(10.5,3.9)$ and $(12.5,4.5)$, exhibit only marginally worse residuals at high $\beta$.

\section{Generalized Soliton Mass Formula}
\label{sec:g-appendix}

\begin{figure}
    \centering
    \includegraphics[width=\columnwidth]{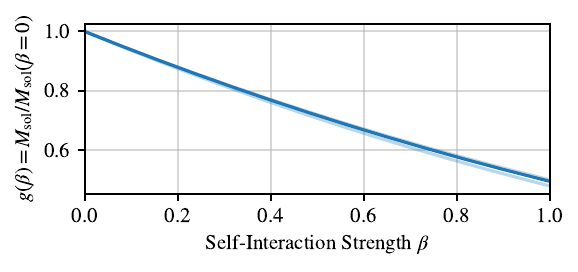}
    \caption{The residual dependence of the soliton mass on self-interaction parameter $\beta$ after normalizing by the zero-SI case. Increasing self-interaction strength decreases the soliton mass relative to the zero-SI prediction. As in Figure \ref{fig:residuals-vs-beta}, the optimal $(a,b)$ parameter set is represented by the bold blue curve, while two other well-fitting parameter sets are shown in faded blue. Changing parameter sets along the degeneracy shown in Figure \ref{fig:residuals-grid} changes $g(\beta)$ only very slightly.}
    \label{fig:soliton-mass-beta-dependence}
\end{figure}

In Eq. \ref{eq:soliton-mass-SI}, we assert that the formula for the mass of dilute solitons under attractive self-interactions is only off by a multiplicative function $g(\beta)$ from the analytic zero-SI formula, Eq. \ref{eq:soliton-mass}. To see this, let $u = r/r_c$ so that
\begin{align*}
    M_{\mathrm{sol}} &= \int_0^{\infty} 4\pi r^2 \rho_{\mathrm{sol}}(r)\mathrm{d}r \\
    &= 4\pi\rho_0 r_c^3 \int_0^{\infty} \frac{u^2 \mathrm{d}u}{\left[1 + A(\beta)u^{B(\beta)} \right]^8}
\end{align*}
where $A(\beta)=0.091a^{\beta/b}$ and $B(\beta)=2-\beta/b$. In this form, it is clear that $M_{\mathrm{sol}} \propto \rho_0 r_c^3 \propto r_c^{-1}$ regardless of SI strength. The multiplicative function $g(\beta)$ is proportional to the above integral,
\begin{align*}
    g(\beta) = \frac{M_{\mathrm{sol}}(\rho_0,\beta)}{M_{\mathrm{sol}}(\rho_0,0)} = \frac{4096A(0)^{3/2}}{33\pi} \int_0^{\infty} \frac{u^2 \mathrm{d}u}{\left[1 + A(\beta)u^{B(\beta)} \right]^8}
\end{align*}
which cannot be expressed in terms of elementary functions except in special cases. A plot of $g(\beta)$, computed with numerical integration, is shown in Figure \ref{fig:soliton-mass-beta-dependence} for different $(a,b)$ parameter sets. The soliton mass under strong self-interactions can be less than $60\%$ of the mass predicted from the zero-SI formula.


\bibliographystyle{mnras}
\bibliography{main_sifdm.bbl}


\bsp	
\label{lastpage}
\end{document}